\documentclass[fleqn,usenatbib]{mnras}

\usepackage{newtxtext,newtxmath}

\usepackage[T1]{fontenc}
\usepackage{ae,aecompl}

\usepackage{graphicx}	
\usepackage{amsmath}	
\usepackage{multirow}   
\usepackage{color}

\title[A search for auroral radio emission from $\beta$ Pictoris b]{A search for auroral radio emission from $\beta$ Pictoris b}

\author[Y. Shiohira et al.]{
Yuta Shiohira$^{1}$\thanks{E-mail: 210d9022@st.kumamoto-u.ac.jp},
Yuka Fujii$^{2}$,
Hajime Kita$^{3}$,
Tomoki Kimura$^{4}$
Yuka Terada$^{5,6}$,
\newauthor
and Keitaro Takahashi$^{2,7}$
\\
$^{1}$Graduate School of Science and Technology, Kumamoto University, Kumamoto, Japan\\
$^{2}$National Astronomical Observatory of Japan, Tokyo, Japan\\
$^{3}$Tohoku Institute of Technology, Sendai, Japan\\
$^{4}$Tokyo University of Science, Tokyo, Japan\\
$^{5}$Institute of Astronomy and Astrophysics, Academia Sinica, Taipei, Taiwan, R.O.C.\\
$^{6}$Department of Astrophysics, National Taiwan University, Taipei, Taiwan, R.O.C.\\
$^{7}$International Research Organization for Advanced Science and Technology, Kumamoto University, Kumamoto, Japan
}

\date{Accepted 2023 December 22. Received 2023 December 22; in original form 2023 August 07}

\pubyear{2023}

\begin{document}
\label{firstpage}
\pagerange{\pageref{firstpage}--\pageref{lastpage}}
\maketitle

\begin{abstract}
Magnetized exoplanets can serve as the source of auroral radio emissions, allowing us to characterize the magnetospheric properties of these planets. Successful detections of auroral radio emissions from brown dwarfs, as well as from Jupiter, suggest that Jupiter-like planets in distant orbits may also generate radio emissions through a similar mechanism.
In this study, we present our search for 250-500 MHz emissions from $\beta$ Pictoris b, one of the most extensively studied young Jupiter-like planets. We conducted the search using the upgraded Giant Metrewave Radio Telescope (uGMRT). Despite the favourable orbital inclination, no signal was detected, putting 3$\sigma$ upper limits on the radiation at 0.18 mJy.
We translate this limit into constraints on the ionospheric and magnetospheric parameters, assuming that the emission is powered by the Hill current system. While the upper limit is larger by a factor of a few than the nominal estimate of radio intensity, we put constraints on the magnetospheric and ionospheric parameters. 
\end{abstract}




\section{Introduction}

Magnetized planets in the solar system generate intense auroral radio emission via the electron cyclotron maser instability (ECMI), driven by the interaction between the planetary magnetic field and the energetic ($\sim\mathrm{keV}$) plasmas in the planetary magnetosphere \citep{wulee1979, zarka1998, treumann2006}.
Similarly, magnetized exoplanets are expected to generate auroral radio emissions. 
Once it is detected, the auroral radio emissions can constrain the magnetic field strengths of the exoplanets model-independently, 
as the maximum frequency of auroral radio emission, $\nu_\mathrm{ce,max}$, is proportional to the strength of the planetary surface magnetic field 
at the poles, $B_\mathrm{pol,p}$ in a non-relativistic regime:
\begin{eqnarray}
    \centering
    \label{freq}
    \nu_\mathrm{ce} \sim 2.8\ \mathrm{MHz} \times B_\mathrm{pol,p} [\mathrm{G}] .
\end{eqnarray}
In addition, the auroral emission can also allow us to probe some properties of the planet, such as the rotation period of the planetary interior (the dynamo region) \citep[e.g.,][]{sanchez-Lavega2004, reiners2010, lazio2019}, the tilt of the magnetic axis with respect to the spin rotation axis \citep{hess2011}, and the plasma source flux. 
Therefore, low-frequency radio observation can provide unique access to exoplanets complementary to the observations by ultraviolet, optical, and infrared observations. 

Detecting auroral emission from exoplanets has been envisioned since before the first detection of an exoplanet \citep[e.g.,][]{yantis1977, winglee1986}. 
The early studies found the empirical law between incident kinetic or magnetic energy of solar wind into the solar system planet's magnetosphere and auroral radio power, ``Radiometric Bode's Law'', which encouraged subsequent studies \citep{deschkaiser1984, zarka1992}. 
Extrapolation from this law suggested that hot Jupiters would generate $10^3-10^5$ times stronger auroral radio emission than Jupiter's, as they are exposed to extreme stellar wind \citep[e.g.,][]{farrell1999, zarka2001, lazio2004} and/or coronal mass ejection \citep{griessmeier2007}. 
Candidates for the intense auroral radio source were also explored from the viewpoint of the properties of the host star wind, taking account of the observed data \citep{stevens2005}, the dependence on the stellar age \citep{griessmeier2005,griessmeier2007} and on the spectral type of the host star \citep{katarzynski2016}, and the evolution after the main-sequence stars \citep{fujii2016}. 

Motivated by these theoretical studies, a number of observations targeted at hot Jupiters and other planets that are likely to receive intense stellar wind \citep{zarka1997, zarka2011, bastian2000, bastian2018, farrell2003, lazio2004, lazio2007, lazio2010a, lazio2010b, ryabov2004, majid2006, shiratori2006, winterhalter2006, george2007, lecavelier2009, lecavelier2011, lecavelier2013, smith2009, stroe2012, hallinan2013, Murphy2015, bower2016, lynch2017, lynch2018, turner2017, o'gorman2018, deGasperin2020, green2021, narang2021b, narang2021a}. 
Although there is a recent claim of a tentative detection of a signal \citep{turner2021}, no univocal detection has been made yet.
This is puzzling given that the highest sensitivities that these observations achieved appear to be sufficient to detect the most intensive radio emission strength predicted, which exceeds tens of mJy at low-frequency \citep[e.g.,][]{griessmeier2017} .
A possible reason argued so far is that the emission beam does not direct to the Earth. 
However, Jupiter-like solid angle of emission, 1.6 str, would correspond to as much as 13\% chance of detection, and it is not likely to fully explain that all of several tens of observations yielded no clear detection.
Therefore, there are other factors that prevent canonical hot Jupiters from emitting intense auroral radio wave at observable frequency range. 
Indeed, recent work suggested that the expansion of the highly ionized upper atmospheres of hot Jupiters may hinder the generation of auroral emission due to the high local plasma frequency \citep{daley-yates2017, daley-yates2018, weber2017a, weber2017b, weber2018}. 
In addition, the fact that hot Jupiters are likely tidally locked to a synchronously rotating state might also suppress the emission by weakening the magnetic field strength \citep{griessmeier2004}. Overall, these considerations and the lack of clear detection so far may indicate that the close-in planetary systems may not be as promising targets as initially thought.

Meanwhile, the auroral emission may also be driven by the coupling between the magnetosphere and the ionosphere of the planet. 
A large fraction of Jovian auroral emission is powered by the current system connecting the magnetosphere and the ionosphere, which is driven by the interaction among Jovian planetary magnetic field ($\sim 10$ G), the fast rotation ($\sim 10$ h), and the plasma particles supplied by Io's volcanic activity \citep{hill1979}. 
By analogy, \cite{nichols2011, nichols2012} proposed that distant giant exoplanets with fast rotation, strong magnetic field and the supply of plasma into the magnetosphere would be suitable targets for radio observation. 
They also pointed out that the XUV intensity of the host star is also a key factor as high XUV irradiation onto the planet increases the ionospheric conductivity, resulting in a larger current. 
Exoplanets may also generate auroral emission by interacting with exomoons \citep{noyola2014, noyola2016}. \cite{narang2023a} conducted the first search for signals from three exoplanets speculated to have Io-like exomoons with Giant Meterwave Radio Telescope (GMRT) and upgrade Giant Meterwave Radio Telescope (uGMRT). No signals were detected from the system on any exoplanetary systems, and the work put upper limits 0.18-1.6 mJy at 150, 325 and 400 MHz. In \cite{narang2023b}, another candidate which also may have volcanic exomoon was targeted by uGMRT observation. Still, the signals were undetected and put upper limits of 0.9-3.3 mJy at 150 MHz and 218 MHz.
These auroral emission mechanisms do not require the planet to be close to the host star. Exoplanets in distant orbits can also be a promising target of low-frequency radio observations.

Interestingly, auroral emission has been successfully detected from brown dwarfs \citep{berger2001, berger2005, berger2002, berger2006, berger2009, burgasser2005, burgasser2013, burgasser2015, osten2006, osten2009, hallinan2007, hallinan2008, hallinan2015, phan-bao2007,  mclean2011, mclean2012, antonova2013, williams2013, gizis2016, lynch2016, kao2016, kao2018, kao2022, route2016, guirado2018, richey-yowell2020, hughes2021, climent2022, vedantham2023} and recently from low-mass stars \citep{vedantham2020, perez-torres2021, callingham2021, pineda2023}. 
Although the exact mechanism has not been elucidated, these emissions may be generated through the current system similar to that in the Jovian magnetosphere. 
Indeed, low-mass stars and brown dwarfs lie between Jovian planets and the solar-type stars in terms of mass and internal structure \citep{chabrier2000} and may develop similar magnetospheric structures. 
If they have a small companion, the system of the star (or brown dwarf) and the companion would be a scaled-up version of the Jupiter-Io system \citep[e.g.,][]{zarka2007, saur2013, turnpenney2018}. 

These observations also motivate one to search for the auroral emission from the young massive Jovian planets, as they have even more similarity to brown dwarfs in that they have large heat flux from the interior and that they are more massive than Jupiter. 
According to one of the proposed scaling laws, these properties suggest the possibility of large magnetic field strengths of young Jovian planets (see Section 2.1 below). 
Large magnetic field strengths are also supported by the observations of gas accretion onto a young planet \citep{hasegawa2021}. 

Along this line, \cite{cendes2022} conducted GHz frequency radio observations of five directly imaged young Jovian planets (Ross 458, GU Psc, 51 Eri, GJ~504 and HR 8799), with Very Large Array (VLA), and put upper limits at 6-210 $\mu $Jy level.
\cite{narang2022} analyzed observational data at frequencies from 150 MHz to 3 GHz obtained from GMRT and VLA to search for radio emission from the directly imaged exoplanet 1RXS1609 b and put upper limits 0.21-6 mJy. 
More survey is necessary to reveal the nature of the possible emissions from young Jupiter-like planets. 

In this paper, we present a search for the auroral radio emission in the 250-500 MHz bandpass from one of the best-studied exoplanets $\beta$ Pictoris b. 
A unique aspect of this target among all the directly-imaged planets is its nearly edge-on orbit. 
It is known that the most vigorous Jupiter's auroral emission is observed from the equatorial plane due to the anisotropic nature of the radio emission \citep{ladreiter1989}. A similar anisotropy implies that the emission is highest if the planet is in an edge-on orbit, on the assumption that the obliquity of the planet (the angle between the orbital axis and the spin axis) is slight. 
$\beta $ Pictoris b also has the measured rotation velocity, which suggests the rotation period of $\sim 8$ hours \citep{snellen2014}. 
These factors would work favourably to observe the intense radio emission from this target. 

The organization of this paper is as follows:
Section 2 presents the properties of $\beta$ Pictoris b and the estimation of the maximum frequency and flux density of the emission. We describe observations and data analysis in Section 3 and report the results in Section 4. 
Lastly, we discuss the constraints on the $\beta$ Pictoris b parameters and summarize this paper in Section 5.

\section{Target}
\label{s2}

The target planetary system in our observation is $\beta$ Pictoris system, which is 19.75 pc away from the Earth. In this system, an A6 V star \citep{gray2006} hosts two exoplanets,  $\beta$ Pic b and c. Both of them are in long orbits and are massive ($\beta$ Pic b: $a=10.2^{+0.4}_{-0.3}\ \mathrm{AU}$, $M_\mathrm{p} =12.8^{+5.3}_{-3.2}\ M_\mathrm{J}$, $\beta$ Pic c: $a=2.68\pm 0.02\ \mathrm{AU}$, $M_\mathrm{p}=8.89\pm 0.75\ M_\mathrm{J}$ \citep{nielsen2020,lacour2021} ). This system is young, 22 Myr old \citep{mamajek2014}, and the planet b has high effective temperature, $T_{\rm eff}=1724\pm 15\ \mathrm{K}$ \citep{chilcote2017} due to the remnant accretion energy \citep{chilcote2017,nowak2020}. 

The possible advantages of the systems in terms of the detectability of auroral radio emission are three-fold.  
First, according to the scaling law for the strength of the planetary magnetic field, planets which are massive and have high luminosity including $\beta $ Pic b tend to have strong magnetic field. 
The maximum frequency of auroral emission is therefore expected to be high, potentially overlapping the GMRT bands with high sensitivity. 
Second, the rotation period of $\beta $ Pic b is estimated to be $\sim 8$~hr, based on the high-resolution spectra \citep{snellen2014}, comparable to Jovian rotation period. 
The short rotation period would be necessary to drive strong auroral emission through magnetosphere-ionosphere (M-I) coupling similar to Jovian auroral emission. 
The estimated rotation period would also help us distinguish the auroral radio emission of the planet from other sources. 
Third, the planets have nearly edge-on orbit ($i= 88.88 \pm 0.09 \ \rm deg$ \citep{nielsen2020}). 
Jupiter's auroral radio emission is detectable only from near the equatorial plane due to the beaming effect \citep{ladreiter1989} (approximately $\pm $10 degree from the  equator), and assuming the similar beaming and the small obliquity make planets on nearly edge-on orbits the only targets whose auroral emissions are detectable from Earth. 
These properties let us select $\beta$ Pic b as the primary target of the search of the auroral radio emission from directly-imaged planets. 

In the following, we describe how we estimate the frequency and flux density of the auroral radio emission of $\beta$ Pic b.

\begin{table*}
    \label{properties}
    \centering
    \begin{tabular}{l|l|l|l}
    
    \hline
               & Parameters & Values & Reference \\
    \hline
    \hline
    
    \multirow{4}{*}{System}     
               & Distance   & 19.75 pc           
               & \multirow{2}{*}{Gaia DR2} \\
               
               & RA         & 05h 47m 17.1s      
               & \multirow{2}{*}{\citep{GaiaCollaboration2016, GaiaCollaboration2018}}\\
               
               & Dec        & $-51^{\circ}$ 03' 59"  
               & \\
               
               & Age        & 22 Myr             
               & \citet{mamajek2014} \\
    \hline

    \multirow{2}{*}{Host star} 
               & Spectral type & A6 V  
               & \citet{gray2006} \\
               
               & Mass          & 1.77(±0.03) $\mathrm{M_{\odot}}$
               & \citet{nielsen2020} \\
    \hline

    \multirow{8}{*}{Planet ($\beta$  Pic b)}
                   
               & Mass & $12.8^{+5.3}_{-3.2}$ $\mathrm{M_J}$
               & \citet{nielsen2020} \\
               
               & Radius        & 1.46($\pm$0.01) $\mathrm{R_J}$
               & \citet{chilcote2017} \\
               
               & Effective temperature  & 1,724($\pm$15) K
               & \citet{chilcote2017} \\
               
               & Semi-major axis  & $10.2^{+0.4}_{-0.3}$ AU
               & \citet{nielsen2020} \\
               
               & Orbital period   & $24.3^{+1.5}_{-1.0}$ yrs
               & \citet{nielsen2020} \\
               
               & Orbital inclination  & 88.88($\pm$0.09) deg
               & \citet{nielsen2020} \\
               
               & Spin velocity  & 25($\pm$3) km/s
               &\multirow{2}{*}{\citet{snellen2014}} \\
               
               & Spin period    & 8 hr (assuming zero obliquity)
               &
               \\
    \hline
    \end{tabular}
    
    \caption{The parameters for $\beta$ Pic system, host star and $\beta$ Pic b}
\end{table*}

\subsection{Maximum frequency of radio emission}
\label{freq_est}

As described in Sec. 1, the maximum frequency of the auroral radio emission $\nu_\mathrm{ce,max}$ depends on the polar magnetic field $B_\mathrm{pol,p}$, given as Eq. \ref{freq}. 
To predict the maximum frequency from that equation, it is necessary to estimate $B_\mathrm{pol,p}$. 
Several scaling laws for the strength of planetary magnetic field have been proposed so far. 
In this study, we consider the law of \cite{christensen2009} and of magnetic Bode's law \citep[e.g.,][]{blackett1947, russell1978, farrell1999}. 
The main difference between the two is the dependence on the planetary rotation. 

Christensen's scaling law states that the strength of the magnetic field on the surface of core region, $B_\mathrm{c,p}$, is independent of the planetary rotation and estimated by the following,
\begin{equation}
    \label{christensen_est}
        B_\mathrm{c,p} ^2
    \propto f_\mathrm{ohm} {\rho_\mathrm{c,p}}^{1/3}  (Fq_{\mathrm{o,p}})^{2/3}
\end{equation}
where $f_\mathrm{ohm}(\leq 1)$ is the ratio of the ohmic dissipation to the total dissipation, $\rho_\mathrm{c,p}$ is the mean density of the core, $F$ is an efficiency factor and of order unity and $q_\mathrm{o,p}$ is the heat flux on the outer surface of the core. To estimate $B_\mathrm{c,p}$, it is necessary to evaluate $\rho_\mathrm{c,p}$ and $q_\mathrm{o,p}$. 
We estimated the $\rho_\mathrm{c,p}$ to be $\sim 7.254\ \mathrm{g/cm^3}$ by calculating the density profile inside the planet assuming a polytropic gas sphere with the index of 1.5, and by determining the radius of the outer boundary of the core to be $r_\mathrm{c,p}\sim 0.9\ R_p$ based on the density that hydrogen undergoes a phase transition to metallic, $\sim 0.7\ \mathrm{g/cm^3}$\citep{griessmeier2007}(see Appendix). 
The $q_\mathrm{o,p}$ is estimated from the assumption that the luminosity of the planet is equal to the total energy flux at the surface of the core. 
Thus, denoting the effective temperature of the planet by $T_\mathrm{eff,p}$, 
\begin{equation}
    \label{heat_flux}
    q_\mathrm{o,p} = \sigma_\mathrm{sb} T_\mathrm{eff,p}^4 \times \left(\frac{R_\mathrm{p}}{r_\mathrm{c,p}}\right)^2
\end{equation}
where $\sigma_\mathrm{sb} \sim 5.67\times10^{-8}\mathrm{Wm^{-2}K}$ is the Stefan-Boltzmann constant. We relate $B_\mathrm{pol,p}$ with $B_\mathrm{c,p}$ by
\begin{equation}
    B_\mathrm{pol,p} \propto B_\mathrm{c,p} \left(\frac{r_\mathrm{c,p}}{R_\mathrm{p}}\right)^3
                 \propto {\rho_\mathrm{c,p}}^{1/6}{q_\mathrm{o,p}}^{1/3}\left(\frac{r_\mathrm{c,p}}{R_\mathrm{p}}\right)^3
\end{equation}
to find the scaling of the surface magnetic field strength as follows:
\begin{equation}
    \label{mag_p}
    B_\mathrm{pol,p} = 
    \left(\frac{\rho_\mathrm{c,p}}{\rho_\mathrm{c,J}} \right)^{1/6}
    \left(\frac{q_\mathrm{o,p}}{q_\mathrm{o,J}} \right)^{1/3}
    \left(\frac{r_\mathrm{c,p}}{r_\mathrm{c,J}} \right)^3
    \left(\frac{R_\mathrm{J}}{R_\mathrm{p}} \right)^3
    B_\mathrm{pol,J}
\end{equation}
The subscript $J$ indicates the Jovian values of the parameters: $\rho_\mathrm{c,J} \sim 1.899\ \mathrm{g/cm^3}$, $R_\mathrm{J} \sim 7.0\times10^7\ \mathrm{m}$, $r_\mathrm{c,J}\sim 0.85\ R_\mathrm{J}$, and $q_\mathrm{o,J} \sim 7.485 \mathrm{W/m^2}$ \citep{li2018}. 
Putting all the numbers above into Eq. \ref{mag_p} gives  $B_\mathrm{pol,p} \sim 66\ B_\mathrm{pol,J}$. 
Therefore, the maximum frequency of auroral radio emission from $\beta$ Pic b would be 66 times larger than that of Jupiter, about 1.8 GHz in this case.

Next, we estimate the frequency from the magnetic Bode's law, which suggests the proportionality between the angular momentum and magnetic moment of a planet:
\begin{equation}
    \label{mag_bode_law}
    B_\mathrm{eq,p}R_\mathrm{p}^3 
    \propto \omega_\mathrm{p} M_\mathrm{p} R_\mathrm{p}^2
\end{equation}
where $B_\mathrm{eq,p}$ is the surface magnetic field strength at the equator, and $\omega_\mathrm{p}\ (\sim 1.25\ \omega_\mathrm{J})$ is the angular velocity of planetary rotation. Therefore, the surface magnetic field strength of $\beta$ Pic b is expressed as follows, 

\begin{equation}
    \label{mag_bode_law_est}
    B_\mathrm{pol,p} =  
    \left(\frac{\omega_\mathrm{p}}{\omega_\mathrm{J}} \right)
    \left(\frac{M_\mathrm{p}}{M_\mathrm{J}} \right)
    \left(\frac{R_\mathrm{p}}{R_\mathrm{J}} \right)^{-1} 
    B_\mathrm{pol,J}
\end{equation}
where we assumed that the planetary magnetic field is a dipole filed, that is $B_\mathrm{pol,p}=2B_\mathrm{eq,p}$. The surface field strength at the polar of $\beta$ Pic b is thus expected to reach $\sim 11\ B_\mathrm{pol,J}$. In this case, the maximum frequency is expected to be about 300 MHz. 

In both cases, the maximum of the emission would be higher than 300 MHz and expected to be observable with uGMRT in band 3 (250-500 MHz).

\subsection{Radio flux density}

The main source of the auroral radio emission from $\beta$ Pic b is considered to be the M-I coupling due to the fast rotation and the strong magnetic field estimated earlier. Assuming that the planetary magnetic field is a dipole field, the total power dissipated by the current system of M-I coupling, $P_{\mathrm{0,J}}$, is given by \cite{hill2001} as follows:
\begin{equation}
    \label{dissipation_power_j}
    P_{\mathrm{0,J}} = \frac{2\pi \Sigma_\mathrm{J} B_\mathrm{eq,J}^2 \omega_\mathrm{J}^2 R_\mathrm{J}^4}{L_\mathrm{J}^2},
\end{equation}
where $\Sigma_\mathrm{J}$ is the height-integrated Pedersen conductivity in the polar ionosphere. $L_\mathrm{J}$ is the scale length representing the region in the equatorial plane of the planet where the magnetospheric plasma is co-rotating with the planet's rotation, and is given (in $R_\mathrm{J}$) as follows.
\begin{equation}
    L_\mathrm{J} = \left( \frac{\pi \Sigma_\mathrm{J} B_\mathrm{eq,J}^2 R_\mathrm{J}^2 }
                   {\dot{M}_\mathrm{J}} \right) ^{1/4}
    \label{scale_len}
\end{equation}
where $\dot{M}_\mathrm{J}$ represents the plasma mass outflow rate. In the case of $\Sigma_\mathrm{J}\sim0.6\ \mathrm{mho}$ and $\dot{M}_\mathrm{J}\sim 2000\ \mathrm{kg/s}$, $L_\mathrm{J}$ is about $30\ R_\mathrm{J}$ \citep{hill2001}. In this study, we assume that the power dissipated by the M-I coupling at $\beta$ Pic b, $P_{\mathrm{0,P}}$ can be estimated by scaling each parameter of Eq. \ref{dissipation_power_j} and Eq. \ref{scale_len}. From Eq. \ref{scale_len}, the scale length for $\beta$ Pic b, $L_\mathrm{p}$ is expressed by
\begin{equation}
    L_\mathrm{p} = \left( \frac{\Sigma_\mathrm{p}}{\Sigma_\mathrm{J}} \right)^{1/4}
        \left( \frac{B_\mathrm{eq,p}}{B_\mathrm{eq,J}}  \right)^{1/2}
        \left( \frac{R_\mathrm{p}}{R_\mathrm{J}}  \right)^{1/2}
        \left( \frac{\dot{M}_\mathrm{p}}{\dot{M}_\mathrm{J}}  \right)^{-1/4} L_\mathrm{J}
    \label{scale_len_p}
\end{equation}
Thus, the total power dissipated by the current system of $\beta$ Pic b's M-I coupling, $P_{\mathrm{0,P}}$ is expressed as follows.
\begin{equation}
    P_{\mathrm{0,P}} = \left( \frac{\Sigma_\mathrm{p}}{\Sigma_\mathrm{J}} \right)^{1/2}
        \left( \frac{\omega_\mathrm{p}}{\omega_\mathrm{J}} \right)^{2}
        \left( \frac{R_\mathrm{p}}{R_\mathrm{J}}  \right)^{3}
        \left( \frac{B_\mathrm{eq,p}}{B_\mathrm{eq,J}}  \right)
        \left( \frac{\dot{M}_\mathrm{p}}{\dot{M}_\mathrm{J}}  \right)^{1/2} P_\mathrm{0,J}
    \label{dissipation_power_p}
\end{equation}

Both $\Sigma_\mathrm{p}$ and $\dot{M}_\mathrm{p}$ are uncertain. 
We scaled $\Sigma _{\rm p}$ with the EUV flux at the planetary orbit, based on the observations of the solar system planets indicating that the ionospheric conductivity is correlated with the solar EUV flux that the planet receives \citep[e.g.,][]{robinson1984,tao2010,kimura2013}. Taking account of the instrinsic EUV flux of $\beta $ Pic approximately 1 order of magnitude smaller than the Sun \cite{sanz-forcada2011} as well as the semi-major axis of $\beta$ Pic b, we estimate the conductivity to be  $\Sigma _p = (5.2/10.2)^2\,(1/10)\ \Sigma _J \sim 0.03\ \Sigma_J$.
Assuming $\dot{M}_\mathrm{p}=\dot{M}_\mathrm{J}$ and $B_{\rm eq,P}=66~B_{\rm eq,J}$, the total dissipated power $P_{\mathrm{0,p}}$ would be $\sim  53\ P_{\mathrm{0,J}}$. Assuming further that the ratio of the power of auroral radio emission to the total dissipated power is the same as Jupiter ($\sim $ $10^{-3}$; \cite{zarka2007}), the power of the auroral radio emission from $\beta$ Pic b, $P_{\mathrm{rad,p}}$ would be expected to be $3.3 \times 10^{13}\ \mathrm{W}$, where we adopted the nominal value of the total power dissipated in the Jovian M-I coupling, $P_{\mathrm{0,J}}=6.0\times10^{14}\ \mathrm{W}$ \citep{hill2001}. 

The flux density of auroral radio emission $S_\mathrm{p}$ is calculated using $P_\mathrm{rad,p}$, solid angle $\Omega$, bandwidth of the emission $\Delta \nu$, and distance $d$ from the earth to $\beta$ Pic system:
\begin{equation}
    S_\mathrm{p} = \frac{P_\mathrm{rad,p}}{\Omega d^2 \Delta \nu}.
    \label{flux_density}
\end{equation}
Assuming that the bandwidth of the auroral radio emission is half of the maximum frequency following e.g., \citet{farrell1999}, that the solid angle of the beam is the same as Jupiter's, and $d=19.75\mathrm{pc}$, the estimated the auroral radio emission's flux density is $S_\mathrm{p}\sim 6.0\ \mathrm{\mu Jy}$. 

Note that the frequency and flux estimation of auroral radio emission has ranges depending on the drive model and assumptions for uncertain parameters such as $\rho_\mathrm{c,p}$ and $q_\mathrm{o,p}$. \cite{ashtari2022} also estimated the flux density from $\beta$ Pic b and c but driven by stellar wind-planet interaction, and they have resulted in the emission from the system reaching tens to hundreds of uJy, more than ten times larger than our estimate. 
Noting that the estimated total power of the emission is similar, we find that this discrepancy comes mainly from the difference in the maximum cyclotron frequency where their estimate is smaller than our estimate by two orders of magnitude, reflecting the different assumption for the scaling laws for the magnetic field strength. 
Specifically, our estimate based on the more recent scaling law led to a strong magnetic field strength due to the dependence on the energy flux, while their scaling law did not has that dependence.

\section{Observation and data analysis}
\label{s3}

We observed the $\beta$ Pic system with the uGMRT on 19 June, 11 July, 1 September and 16 September 2020. About 2 hours of observation were carried out at band 3 (centered at 400 MHz with a bandwidh of 200 MHz) each day. 
On every observation day, we conducted the following routine: First, we observed 3C147 for 10 min as a flux and bandpass calibrator. After that, J0538-440, the phase calibrator, and $\beta$ Pic were observed for 5 min and about 30 min, respectively and alternately. In the last 10 min, 3C147 was observed again.

For the reduction of data, we used the pipeline that is CAsa Pipeline-cum-Toolkit for Upgraded Giant Metrewave Radio Telescope data REduction(CAPTURE) \citep{Kale2021}. It executes tasks, flagging, calibration, imaging and self-calibration  utilizing Common Astronomy Software Applications (CASA). 

In the first step, raw visibility data was flagged using CASA tasks \texttt{flagdata}. The frequency channel 0 and first and last 10 s of each scan were flagged. After that, calibration for delay, bandpass and gain were carried out. Flagging and calibration performed again. After this procedure, we use CASA's \texttt{tclean} task for imaging. The size of pixel in these images are $1.0'' \times 1.0''$. After imaging, 4 rounds of phase only self-calibration and 4 rounds of phase and amplitude self-calibration were executed. For final images made through the pipeline, we carried out primary beam correction with tasks \texttt{wbpbgmrt}\footnote{%
\href{https://github.com/ruta-k/uGMRTprimarybeam}{https://github.com/ruta-k/uGMRTprimarybeam}}.
\section{Results}
\label{s4}

We detect no radio emission from $\beta$ Pic b and set 3$\sigma_\mathrm{rms}$ upper limit for each of four observations. 

\begin{figure*}
    \begin{tabular}{cc}
      \begin{minipage}[t]{0.45\hsize}
        \centering
        \includegraphics[width=.8\hsize]{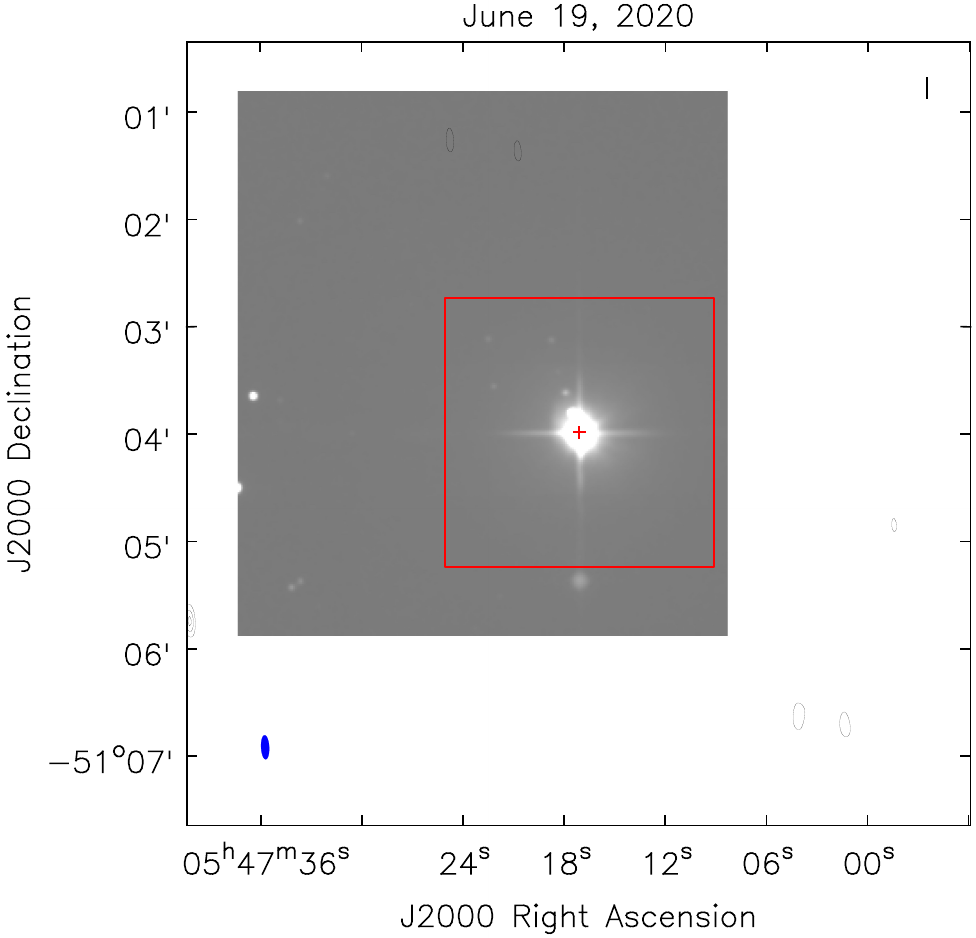}
      \end{minipage} &
      \begin{minipage}[t]{0.45\hsize}
        \centering
        \includegraphics[width=.81\hsize]{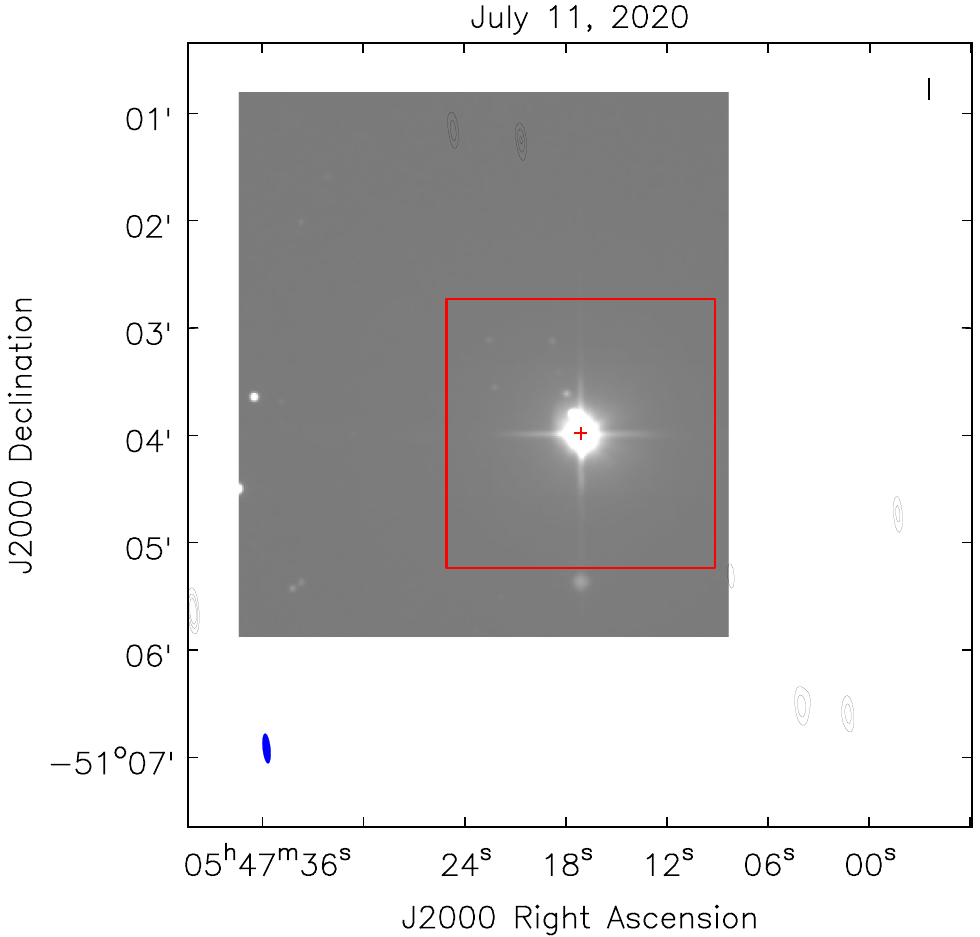}
      \end{minipage} \\
      
      \begin{minipage}[t]{0.45\hsize}
        \centering
        \includegraphics[width=.8\hsize]{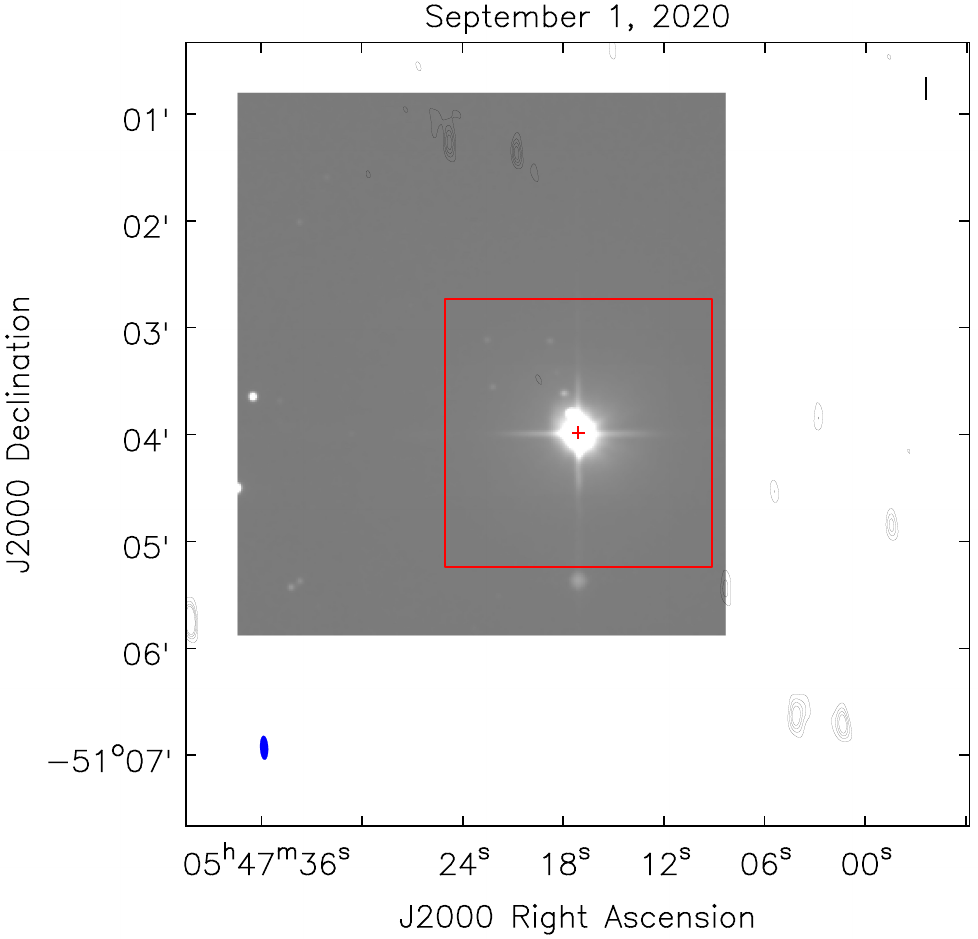}
      \end{minipage} &
      \begin{minipage}[t]{0.45\hsize}
        \centering
        \includegraphics[width=.80\hsize]{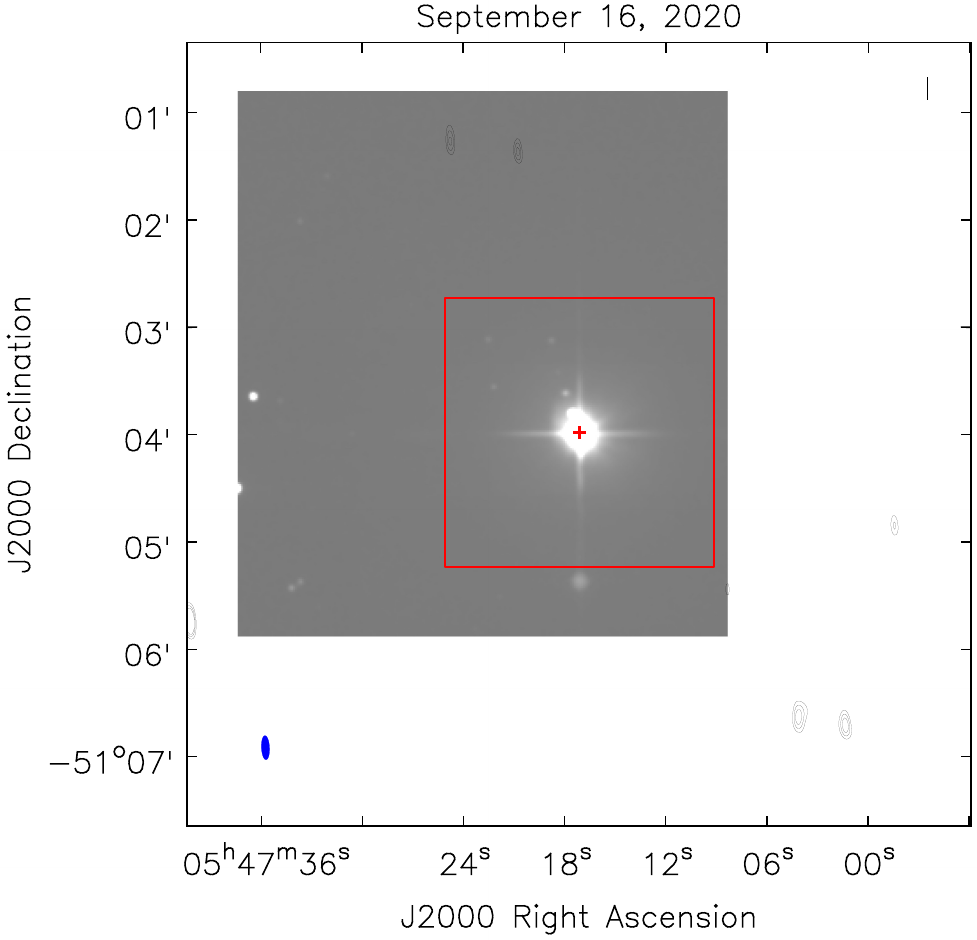}
      \end{minipage} \\
    
    \end{tabular}
    \caption{The uGMRT image for each of the four observation day expressed in contours. All are overlaid on 2MASS infrared image. Contours are set to 5,10,15,20 $\times \sigma_{\mathrm{rms}}$, where $\sigma_{\mathrm{rms}}$ are computed in the area surrounded red solid line,$150'' \times 150''$, and $\sigma_{\mathrm{rms}} = 121\ \mu\mathrm{Jy}\ (6/19), 88\ \mu\mathrm{Jy}\ (7/11), 82\ \mu\mathrm{Jy}\ (9/1), 94\ \mu\mathrm{Jy}\ (9/16)$. The red cross displays the position of the$\beta$ Pic. The blue ellipse shown at left corner express the beam size.}
    \label{images}
    
\end{figure*}

Fig. \ref{images} shows the radio flux contours with the overlaid gray-scale image made by 2MASS All-Sky Data Release at J-band which can be obtained from \textit{Interactive 2MASS Image Service} \footnote{%
\href{https://irsa.ipac.caltech.edu/applications/2MASS/IM/interactive.html}{https://irsa.ipac.caltech.edu/applications/2MASS/IM/interactive.html}}, and the location of the $\beta$ Pic is marked with a red dot in each image.
We computed the root-mean-square (rms) noise of each image on the area where $150'' \times 150''$ centred the target, surrounded by red solid line on the image. The results are listed on Table \ref{rms}, and the contours on Fig.\ref{images} indicate 5,10,15,20 times rms noise.
\begin{table*}
    \centering
    \begin{tabular}{cccc}
    
    \hline
    \multirow{2}{*}{Date}
    & On source time
    & \begin{tabular}{c}
        Synthesized beam \vspace{-1.5mm}\\size
      \end{tabular}
    & rms noise  \\
    
    \quad
    & (hr)
    & (arcsec$\times$arcsec)
    & ($\mu$Jy $\mathrm{beam}^{-1}$)\\
    
    \hline
    \hline
    
    2020 Jun 19
    & 2.3
    & 13.4 $\times$ 4.5 
    & 121 \\
    
    2020 Jul 11
    & 2.0
    & 16.6 $\times$ 4.3
    & 88 \\
    
    2020 Sep 01
    & 1.9
    & 15.0 $\times$ 4.8 
    & 82 \\
    
    2020 Sep 16
    & 1.5
    & 16.0 $\times$ 4.9 
    & 94 \\
    
    (Stacking all)
    & 7.7
    & 14.5 $\times$ 4.7
    & 60 \\
    
    \hline

    \end{tabular}
    
    \caption{Summary of the observations and images.}
    \label{rms}
\end{table*}

The rms noise is substantially larger than the nominal value of band 3 ($\sim 11.5\ \mu\mathrm{Jy}$). 
This is likely due to the low-elevation of the observations. 
When observing an object at low-elevation with interferometre, projected baselines are shortened, and the beam was stretched in Declination direction about two time than the ideal size \citep{gupta2017}. This beam expansion would lead to increase the noise level. 

\begin{figure*}
    \begin{tabular}{cc}
        \begin{minipage}[t]{0.45\hsize}
            \centering
            \includegraphics[width=.75\hsize]{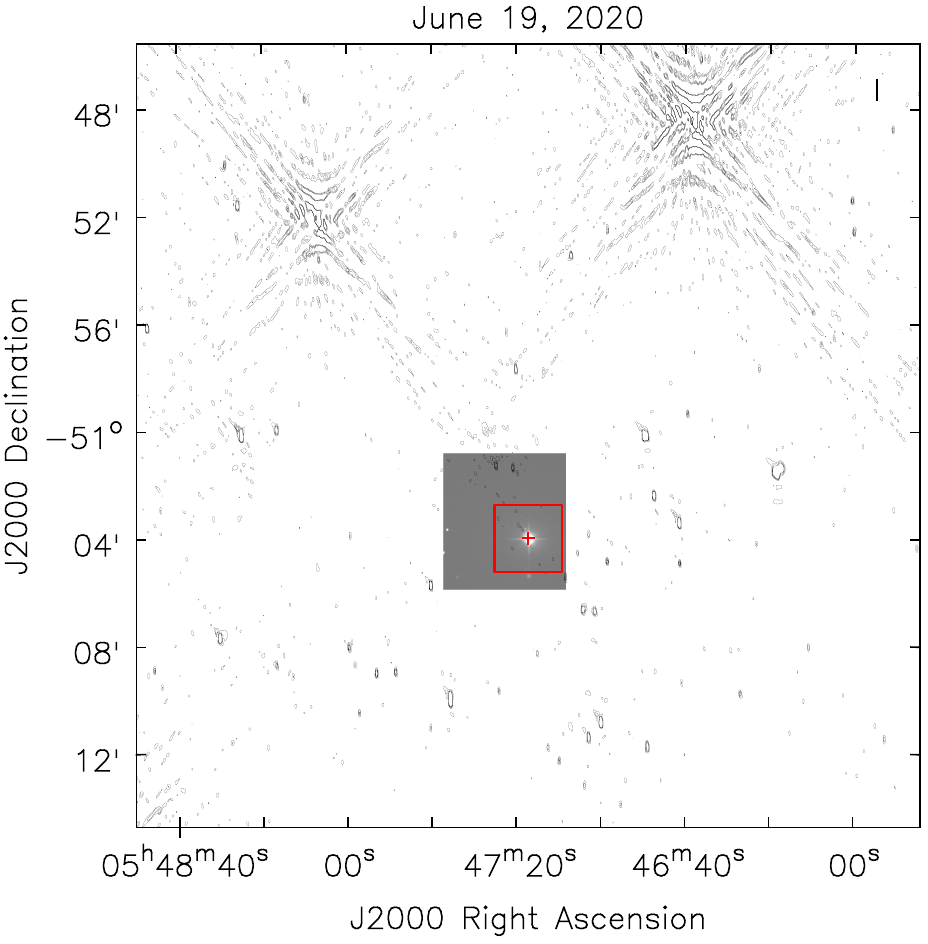}
        \end{minipage} &
        \begin{minipage}[t]{0.45\hsize}
            \centering
            \includegraphics[width=.75\hsize]{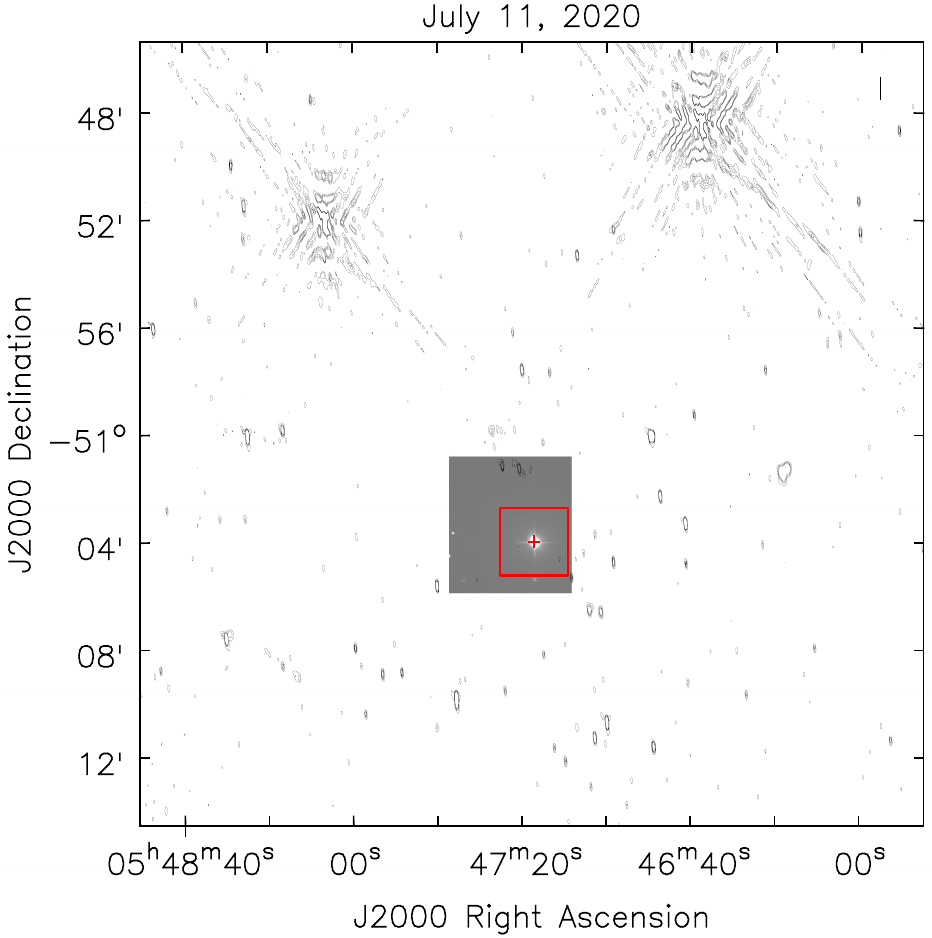}
        \end{minipage}\\
      
        \begin{minipage}[t]{0.45\hsize}
            \centering
            \includegraphics[width=.75\hsize]{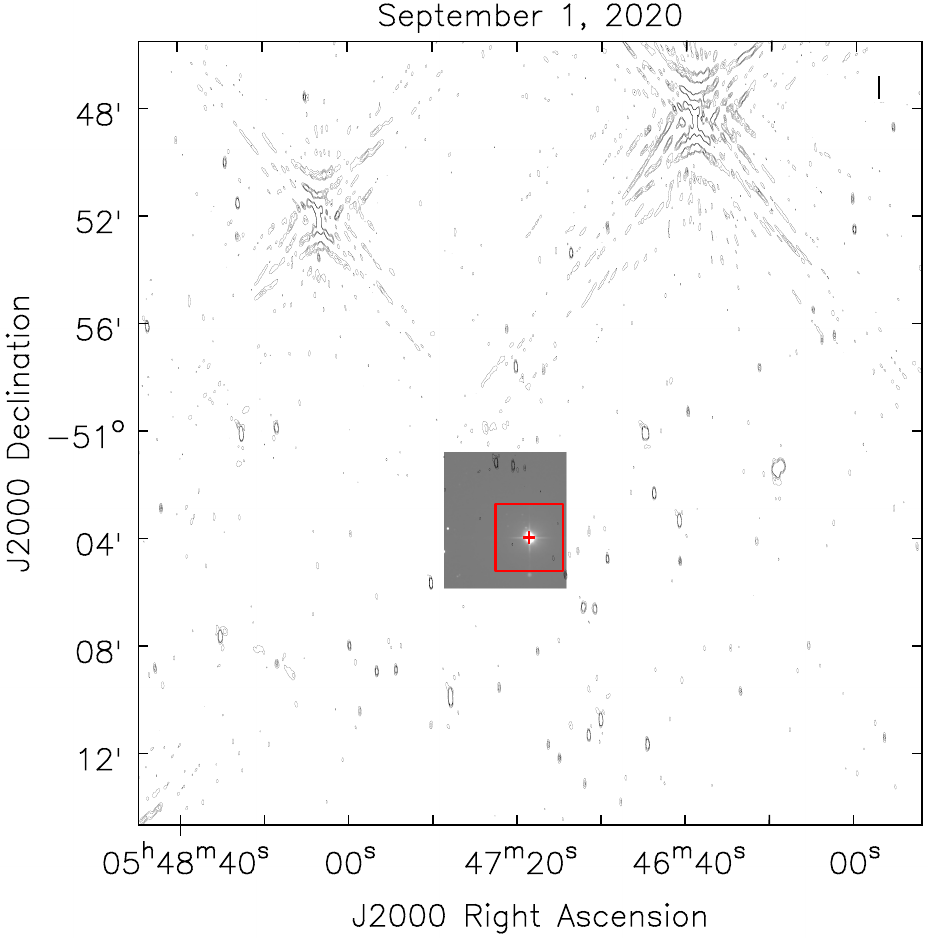}
        \end{minipage} &
        \begin{minipage}[t]{0.45\hsize}
            \centering
            \includegraphics[width=.75\hsize]{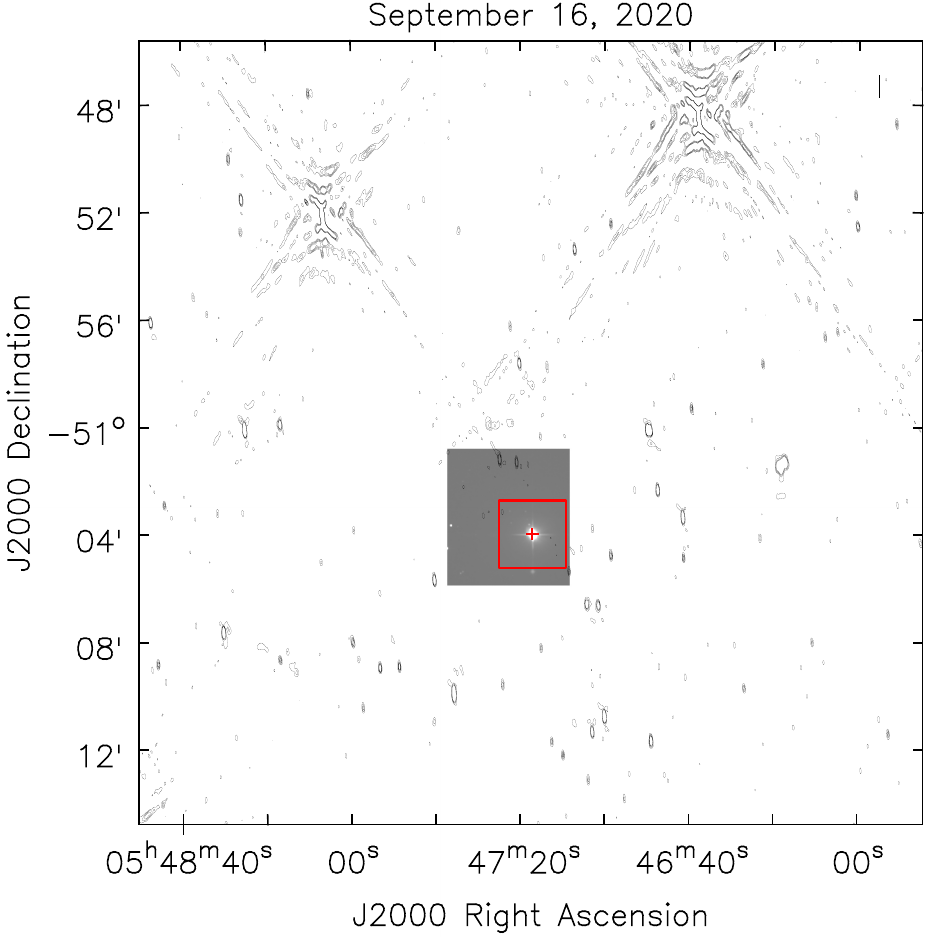}
        \end{minipage}\\
        
    \end{tabular}
    \caption{The uGMRT image for each of the four observation day expressed in contours. Contours are plotted 5,10,15,20 $\times$ 60 $\mu$Jy which value is the quietest rms noise level in the field of view. The red square located near the centre of the image is same area as shown in Fig. \ref{images}. The red cross displays the position of the $\beta$ Pic.}
    \label{sidelobe}
\end{figure*}

In addition, our observation was affected by the side-lobe of intense radio sources which locate near the target. 
Fig. \ref{sidelobe} shows the zoom-out of Fig. \ref{images}, where the contours are now 5, 10, 15 and 20 times the 60 $\mu$Jy. Clearly visible two sources that extend to the position of the target are most likely J054806.4-505206$(\mathrm{RA}=05\mathrm{h} 48\mathrm{m} 7\mathrm{s}, \mathrm{Dec}=-50^{\circ} 52' 7'')$ and J054637.6-504830$(\mathrm{RA}=05\mathrm{h} 46\mathrm{m} 38\mathrm{s}, \mathrm{Dec}-50^{\circ} 48' 28'')$ with flux densities of about 0.1 Jy and 0.2 Jy, respectively. 
While the mrs noise of the background is $\sim $60 $\mu$Jy, that in the side-lobe including the position of our target is more than 80 $\mu$Jy.

\begin{figure*}
    \centering
    \includegraphics[width=.45\hsize]{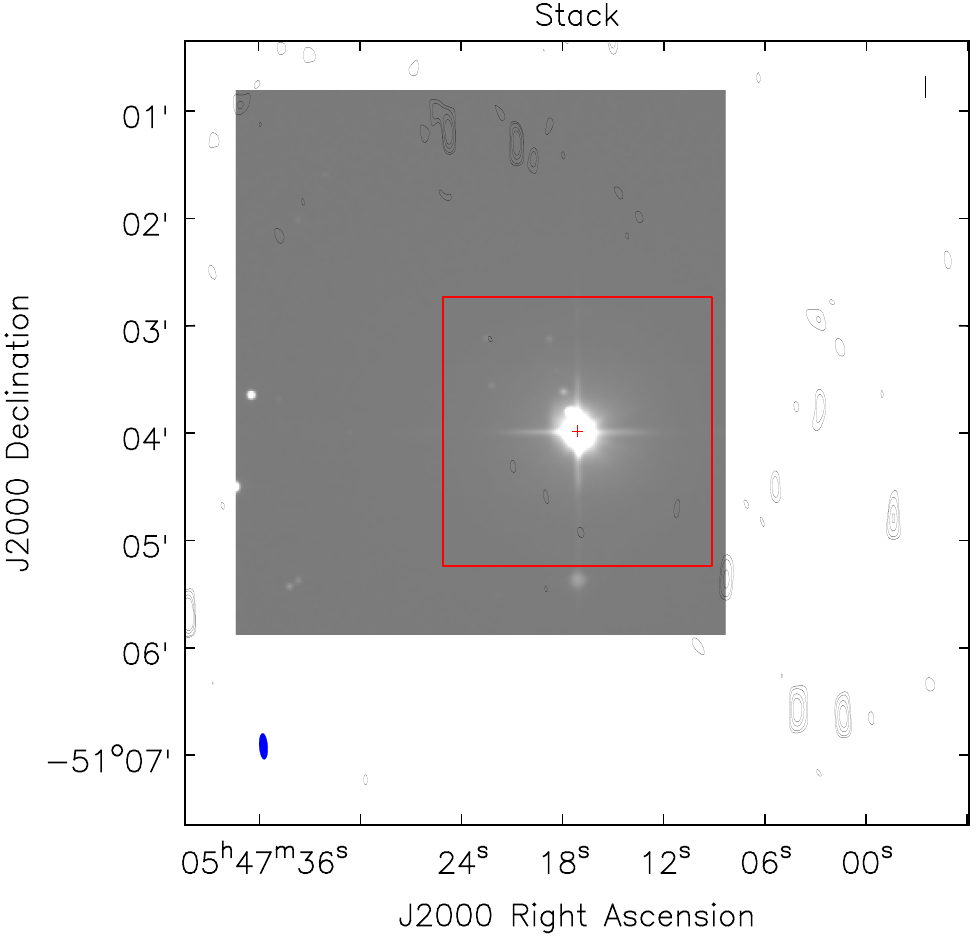}
    \caption{The deep image made by combining the visibility data obtained from all four observations, which is overlaid on 2MASS infrared image. Contours are set to 3,5,10,15 $\times \sigma_{\mathrm{rms}}$, where $\sigma_{\mathrm{rms}}$ are also computed in the area surrounded red solid line, $150'' \times 150''$, and $\sigma_{\mathrm{rms}} = 60\ \mu\mathrm{Jy}$. The red cross displays the position of the $\beta$ Pic. The blue ellipse shown at left corner express the beam size.}
    \label{stack}
\end{figure*}

We made a deeper image combining all four observations, shown in Fig. \ref{stack}, and calculated the rms noise in the same way as Fig. \ref{images} to find 60 $\mu$Jy. This value was 10 times lager than the predicted value of flux density, and no signal was also detected. Then, we set the upper limit of $3\sigma_{\mathrm{rms}} = 180\ \mu\mathrm{Jy}$.

\section{Summary and Discussion}
\label{s5}

In this study, we have reported the results of the observation of the auroral radio emission from $\beta$ Pic b using uGMRT with band 3. We estimated that it could generate a strong magnetic field and intense auroral radio emission through a mechanism similar to Juptier. However, we could not find any signals from the $\beta$ Pic system and put the 3$\sigma $ upper limit of $\sim $ 180 $\mu $Jy. 

While this upper limit is larger than our nominal estimate of the radio emission, we could translate this limit to the constraints on the combination of the planetary parameters, assuming that the planetary radio emission arrives at Earth but was not detected due only to the insufficient sensitivity. 
Let us first consider the case where the planetary magnetic field is stronger than 10~$B_\mathrm{pol,J}$, i.e., the maximum frequency of the emission is within or larger than the bandpass of band 3 of uGMRT. 
Based on the scaling laws introduced in Sec. 2, the emission flux density is given by
\begin{equation}
    \label{sigma_mdot}
    S_\mathrm{p} = \frac{
    \left( \frac{\omega_\mathrm{p}}{\omega_\mathrm{J}} \right)^{2}
    \left( \frac{R_\mathrm{p}}{R_\mathrm{J}}  \right)^{3}
    \alpha_\mathrm{J} P_\mathrm{0,J}}
    {\Omega d^2 (2.8\times10^6\times B_\mathrm{pol,J}[\mathrm{G}])}\  \times\ \frac{\alpha_\mathrm{p}}{\alpha_\mathrm{J}} \left(\frac{\Sigma_\mathrm{p} \dot{M}_\mathrm{p}}{\Sigma_\mathrm{J} \dot{M}_\mathrm{J}}\right)^{1/2}
\end{equation}
where $\alpha$ is the efficiency of the dissipation power to the emission power ($\alpha_\mathrm{J}\sim 10^{-3}$). To apply the upper limit, which was derived from the rms noise of the deepest image, to the parameter constraints, we should consider the total duration of the beam illuminating the Earth $\Delta t'$ and the emission bandwidth occupied within the bandpass $\Delta \nu'$. The upper limit modified for parameter constraints is thus given by 
\begin{equation}
    S_\mathrm{p,lim} = 180[\mu\mathrm{Jy}] \times \left( \frac{\Delta t'}{7.7 \mathrm{hr}}\right)^{-1/2} \left( \frac{\Delta \nu'}{200 \mathrm{MHz}}\right)^{-1/2}~~~~.
\end{equation}
In this paper, we determined $\Delta t'$ by combining the parameters, rotation period  $T_\mathrm{rot}$, the beam solid angle $\Omega$ and initial phase of beam $\phi$. The strength of planetary surface magnetic field controls and $\Delta \nu'$. Substituting $S_\mathrm{p}$ in Eq. \ref{sigma_mdot} by $S_\mathrm{p,lim}$, we obtained the upper limit of the product of $\alpha$, $\Sigma_\mathrm{p}$ and $\dot{M}_\mathrm{p}$.

\begin{figure*}
    \centering
    \includegraphics[width=.6\hsize]{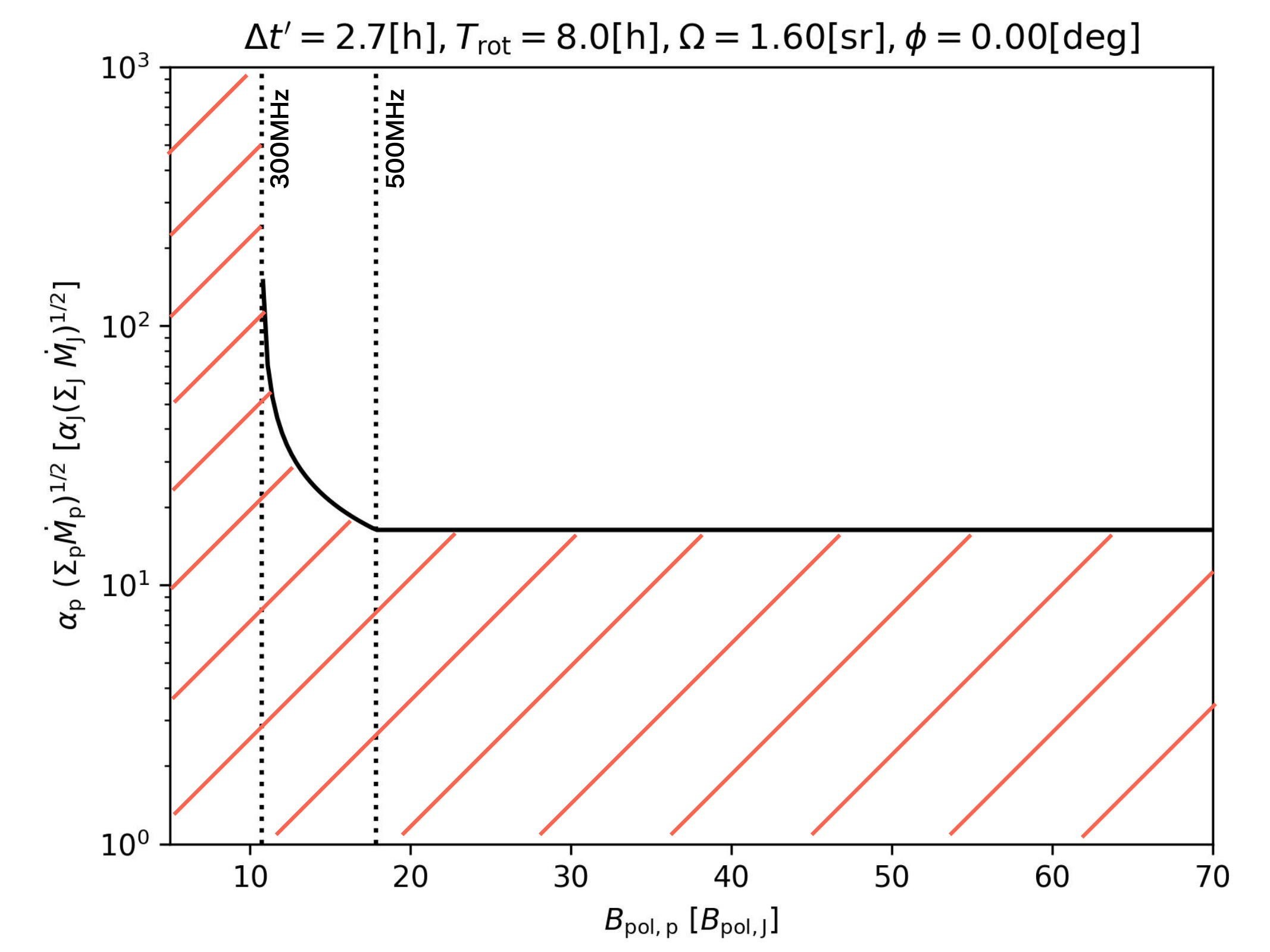}
    \caption{The most strict upper limit of the product of $\alpha_\mathrm{p}$, $\Sigma_\mathrm{p}$, and $\dot{M}\mathrm{p}$ for each magnetic field is derived with the longest observable time $\Delta t' = 2.7 \mathrm{h}$. The range of the product is restricted red shaded area. Two dotted lines illustrate the magnetic field strength which correspond to boundaries of bandwidth, 300 MHz and 500 MHz.}
    \label{lim}
\end{figure*}

If the maximum frequency of the emission is within the bandpass of band 3 (between 250~MHz and 500~MHz), the radio flux detectable in band 3 is reduced and the constraints on the combination of these parameters are weakened accordingly. 
On the other hand, if the maximum frequency is below the observation bandpass, the only constraint may be put on the magnetic field strength. 
The constraints based on these considerations are summarized in Fig. \ref{lim} where the region filled with the diagonal lines are allowed.

We note that there remains the possibility that the emission is not directed toward the Earth. 
This possibility includes not only the case where the beam is always away from the Earth but also the case where the beam happened to be away from the Earth in all of the observation periods. 
If the former is the case, it implies that the obliquity of the planet is larger than Jupiter, or the tilt of the magnetic field is substantial, both of which provide constraints on the formation of Jupiter-like planets. 

In either case, this system would be one of the important targets in the observations with Square Kilometre Array (SKA), which will cover the frequency range similar to our observations with higher sensitivity. 
Because it is located in the Southern hemisphere, it can continuously monitor the target for a longer time and would be less affected by the side-lobes we found, and it will provide a tighter constraint on the combination of planetary parameters to the level of our nominal estimate.

\section*{Acknowledgements}
We thank the staff of the GMRT who have made these observations possible. The GMRT is run by the National Centre for Radio Astrophysics of the Tata Institute of Fundamental Research. Y.S. is supported by JSPS KAKENHI grant No. 21J20760.
Y.F. is supported by a Grant-in-Aid from MEXT of Japan, No. 18K13601. K.T. is partially supported by JSPS KAKENHI Grant Numbers 20H00180, 21H01130 and 21H04467 and the ISM Cooperative Research Program (2023-ISMCRP-2046). H.K. was supported by Grant-in-Aid for JSPS Research Fellow. T.K. is supported by Grants-in-Aid for JSPS KAKENHI Grant Number 19H01948.

\section*{Data Availability}
The observational data used in this paper can be accessed in the GMRT online archive (\href{https://naps.ncra.tifr.res.in/goa/data/search}{https://naps.ncra.tifr.res.in/goa/data/search}) under proposal number 38\_008.




\bibliographystyle{mnras}
\bibliography{ref}

\begin{thebibliography}{}
\makeatletter
\relax
\def\mn@urlcharsother{\let\do\@makeother \do\$\do\&\do\#\do\^\do\_\do\%\do\~}
\def\mn@doi{\begingroup\mn@urlcharsother \@ifnextchar [ {\mn@doi@} {\mn@doi@[]}}
\def\mn@doi@[#1]#2{\def\@tempa{#1}\ifx\@tempa\@empty \href {http://dx.doi.org/#2} {doi:#2}\else \href {http://dx.doi.org/#2} {#1}\fi \endgroup}
\def\mn@eprint#1#2{\mn@eprint@#1:#2::\@nil}
\def\mn@eprint@arXiv#1{\href {http://arxiv.org/abs/#1} {{\tt arXiv:#1}}}
\def\mn@eprint@dblp#1{\href {http://dblp.uni-trier.de/rec/bibtex/#1.xml} {dblp:#1}}
\def\mn@eprint@#1:#2:#3:#4\@nil{\def\@tempa {#1}\def\@tempb {#2}\def\@tempc {#3}\ifx \@tempc \@empty \let \@tempc \@tempb \let \@tempb \@tempa \fi \ifx \@tempb \@empty \def\@tempb {arXiv}\fi \@ifundefined {mn@eprint@\@tempb}{\@tempb:\@tempc}{\expandafter \expandafter \csname mn@eprint@\@tempb\endcsname \expandafter{\@tempc}}}

\bibitem[\protect\citeauthoryear{{Antonova}, {Hallinan}, {Doyle}, {Yu}, {Kuznetsov}, {Metodieva}, {Golden}  \& {Cruz}}{{Antonova} et~al.}{2013}]{antonova2013}
{Antonova} A.,  {Hallinan} G.,  {Doyle} J.~G.,  {Yu} S.,  {Kuznetsov} A.,  {Metodieva} Y.,  {Golden} A.,   {Cruz} K.~L.,  2013, \mn@doi [\aap] {10.1051/0004-6361/201118583}, \href {https://ui.adsabs.harvard.edu/abs/2013A&A...549A.131A} {549, A131}

\bibitem[\protect\citeauthoryear{{Ashtari}, {Sciola}, {Turner}  \& {Stevenson}}{{Ashtari} et~al.}{2022}]{ashtari2022}
{Ashtari} R.,  {Sciola} A.,  {Turner} J.~D.,   {Stevenson} K.,  2022, \mn@doi [\apj] {10.3847/1538-4357/ac92f5}, \href {https://ui.adsabs.harvard.edu/abs/2022ApJ...939...24A} {939, 24}

\bibitem[\protect\citeauthoryear{{Bastian}, {Dulk}  \& {Leblanc}}{{Bastian} et~al.}{2000}]{bastian2000}
{Bastian} T.~S.,  {Dulk} G.~A.,   {Leblanc} Y.,  2000, \mn@doi [\apj] {10.1086/317864}, \href {https://ui.adsabs.harvard.edu/abs/2000ApJ...545.1058B} {545, 1058}

\bibitem[\protect\citeauthoryear{{Bastian}, {Villadsen}, {Maps}, {Hallinan}  \& {Beasley}}{{Bastian} et~al.}{2018}]{bastian2018}
{Bastian} T.~S.,  {Villadsen} J.,  {Maps} A.,  {Hallinan} G.,   {Beasley} A.~J.,  2018, \mn@doi [\apj] {10.3847/1538-4357/aab3cb}, \href {https://ui.adsabs.harvard.edu/abs/2018ApJ...857..133B} {857, 133}

\bibitem[\protect\citeauthoryear{{Berger}}{{Berger}}{2002}]{berger2002}
{Berger} E.,  2002, \mn@doi [\apj] {10.1086/340301}, \href {https://ui.adsabs.harvard.edu/abs/2002ApJ...572..503B} {572, 503}

\bibitem[\protect\citeauthoryear{{Berger}}{{Berger}}{2006}]{berger2006}
{Berger} E.,  2006, \mn@doi [\apj] {10.1086/505787}, \href {https://ui.adsabs.harvard.edu/abs/2006ApJ...648..629B} {648, 629}

\bibitem[\protect\citeauthoryear{{Berger} et~al.,}{{Berger} et~al.}{2001}]{berger2001}
{Berger} E.,  et~al., 2001, \mn@doi [\nat] {10.48550/arXiv.astro-ph/0102301}, \href {https://ui.adsabs.harvard.edu/abs/2001Natur.410..338B} {410, 338}

\bibitem[\protect\citeauthoryear{{Berger} et~al.,}{{Berger} et~al.}{2005}]{berger2005}
{Berger} E.,  et~al., 2005, \mn@doi [\apj] {10.1086/430343}, \href {https://ui.adsabs.harvard.edu/abs/2005ApJ...627..960B} {627, 960}

\bibitem[\protect\citeauthoryear{{Berger} et~al.,}{{Berger} et~al.}{2009}]{berger2009}
{Berger} E.,  et~al., 2009, \mn@doi [\apj] {10.1088/0004-637X/695/1/310}, \href {https://ui.adsabs.harvard.edu/abs/2009ApJ...695..310B} {695, 310}

\bibitem[\protect\citeauthoryear{{Blackett}}{{Blackett}}{1947}]{blackett1947}
{Blackett} P.~M.~S.,  1947, \mn@doi [\nat] {10.1038/159658a0}, \href {https://ui.adsabs.harvard.edu/abs/1947Natur.159..658B} {159, 658}

\bibitem[\protect\citeauthoryear{{Bower}, {Loinard}, {Dzib}, {Galli}, {Ortiz-Le{\'o}n}, {Moutou}  \& {Donati}}{{Bower} et~al.}{2016}]{bower2016}
{Bower} G.~C.,  {Loinard} L.,  {Dzib} S.,  {Galli} P. A.~B.,  {Ortiz-Le{\'o}n} G.~N.,  {Moutou} C.,   {Donati} J.-F.,  2016, \mn@doi [\apj] {10.3847/0004-637X/830/2/107}, \href {https://ui.adsabs.harvard.edu/abs/2016ApJ...830..107B} {830, 107}

\bibitem[\protect\citeauthoryear{{Burgasser} \& {Putman}}{{Burgasser} \& {Putman}}{2005}]{burgasser2005}
{Burgasser} A.~J.,  {Putman} M.~E.,  2005, \mn@doi [\apj] {10.1086/429788}, \href {https://ui.adsabs.harvard.edu/abs/2005ApJ...626..486B} {626, 486}

\bibitem[\protect\citeauthoryear{{Burgasser}, {Melis}, {Zauderer}  \& {Berger}}{{Burgasser} et~al.}{2013}]{burgasser2013}
{Burgasser} A.~J.,  {Melis} C.,  {Zauderer} B.~A.,   {Berger} E.,  2013, \mn@doi [\apjl] {10.1088/2041-8205/762/1/L3}, \href {https://ui.adsabs.harvard.edu/abs/2013ApJ...762L...3B} {762, L3}

\bibitem[\protect\citeauthoryear{{Burgasser}, {Melis}, {Todd}, {Gelino}, {Hallinan}  \& {Bardalez Gagliuffi}}{{Burgasser} et~al.}{2015}]{burgasser2015}
{Burgasser} A.~J.,  {Melis} C.,  {Todd} J.,  {Gelino} C.~R.,  {Hallinan} G.,   {Bardalez Gagliuffi} D.,  2015, \mn@doi [\aj] {10.1088/0004-6256/150/6/180}, \href {https://ui.adsabs.harvard.edu/abs/2015AJ....150..180B} {150, 180}

\bibitem[\protect\citeauthoryear{{Callingham} et~al.,}{{Callingham} et~al.}{2021}]{callingham2021}
{Callingham} J.~R.,  et~al., 2021, \mn@doi [Nature Astronomy] {10.1038/s41550-021-01483-0}, \href {https://ui.adsabs.harvard.edu/abs/2021NatAs...5.1233C} {5, 1233}

\bibitem[\protect\citeauthoryear{{Cendes}, {Williams}  \& {Berger}}{{Cendes} et~al.}{2022}]{cendes2022}
{Cendes} Y.,  {Williams} P.~K.~G.,   {Berger} E.,  2022, \mn@doi [\aj] {10.3847/1538-3881/ac32c8}, \href {https://ui.adsabs.harvard.edu/abs/2022AJ....163...15C} {163, 15}

\bibitem[\protect\citeauthoryear{{Chabrier} \& {Baraffe}}{{Chabrier} \& {Baraffe}}{2000}]{chabrier2000}
{Chabrier} G.,  {Baraffe} I.,  2000, \mn@doi [\araa] {10.1146/annurev.astro.38.1.337}, \href {https://ui.adsabs.harvard.edu/abs/2000ARA&A..38..337C} {38, 337}

\bibitem[\protect\citeauthoryear{{Chilcote} et~al.,}{{Chilcote} et~al.}{2017}]{chilcote2017}
{Chilcote} J.,  et~al., 2017, \mn@doi [\aj] {10.3847/1538-3881/aa63e9}, \href {https://ui.adsabs.harvard.edu/abs/2017AJ....153..182C} {153, 182}

\bibitem[\protect\citeauthoryear{{Christensen}, {Holzwarth}  \& {Reiners}}{{Christensen} et~al.}{2009}]{christensen2009}
{Christensen} U.~R.,  {Holzwarth} V.,   {Reiners} A.,  2009, \mn@doi [\nat] {10.1038/nature07626}, \href {https://ui.adsabs.harvard.edu/abs/2009Natur.457..167C} {457, 167}

\bibitem[\protect\citeauthoryear{{Climent} et~al.,}{{Climent} et~al.}{2022}]{climent2022}
{Climent} J.~B.,  et~al., 2022, \mn@doi [arXiv e-prints] {10.48550/arXiv.2201.12606}, \href {https://ui.adsabs.harvard.edu/abs/2022arXiv220112606C} {p. arXiv:2201.12606}

\bibitem[\protect\citeauthoryear{{Daley-Yates} \& {Stevens}}{{Daley-Yates} \& {Stevens}}{2017}]{daley-yates2017}
{Daley-Yates} S.,  {Stevens} I.~R.,  2017, \mn@doi [Astronomische Nachrichten] {10.1002/asna.201713395}, \href {https://ui.adsabs.harvard.edu/abs/2017AN....338..881D} {338, 881}

\bibitem[\protect\citeauthoryear{{Daley-Yates} \& {Stevens}}{{Daley-Yates} \& {Stevens}}{2018}]{daley-yates2018}
{Daley-Yates} S.,  {Stevens} I.~R.,  2018, \mn@doi [\mnras] {10.1093/mnras/sty1652}, \href {https://ui.adsabs.harvard.edu/abs/2018MNRAS.479.1194D} {479, 1194}

\bibitem[\protect\citeauthoryear{{Desch} \& {Kaiser}}{{Desch} \& {Kaiser}}{1984}]{deschkaiser1984}
{Desch} M.~D.,  {Kaiser} M.~L.,  1984, \mn@doi [\nat] {10.1038/310755a0}, \href {https://ui.adsabs.harvard.edu/abs/1984Natur.310..755D} {310, 755}

\bibitem[\protect\citeauthoryear{{Farrell}, {Desch}  \& {Zarka}}{{Farrell} et~al.}{1999}]{farrell1999}
{Farrell} W.~M.,  {Desch} M.~D.,   {Zarka} P.,  1999, \mn@doi [\jgr] {10.1029/1998JE900050}, \href {https://ui.adsabs.harvard.edu/abs/1999JGR...10414025F} {104, 14025}

\bibitem[\protect\citeauthoryear{{Farrell}, {Desch}, {Lazio}, {Bastian}  \& {Zarka}}{{Farrell} et~al.}{2003}]{farrell2003}
{Farrell} W.~M.,  {Desch} M.~D.,  {Lazio} T.~J.,  {Bastian} T.,   {Zarka} P.,  2003, in {Deming} D.,  {Seager} S.,  eds,  Astronomical Society of the Pacific Conference Series Vol. 294, Scientific Frontiers in Research on Extrasolar Planets. pp 151--156

\bibitem[\protect\citeauthoryear{{Fujii}, {Spiegel}, {Mroczkowski}, {Nordhaus}, {Zimmerman}, {Parsons}, {Mirbabayi}  \& {Madhusudhan}}{{Fujii} et~al.}{2016}]{fujii2016}
{Fujii} Y.,  {Spiegel} D.~S.,  {Mroczkowski} T.,  {Nordhaus} J.,  {Zimmerman} N.~T.,  {Parsons} A.~R.,  {Mirbabayi} M.,   {Madhusudhan} N.,  2016, \mn@doi [\apj] {10.3847/0004-637X/820/2/122}, \href {https://ui.adsabs.harvard.edu/abs/2016ApJ...820..122F} {820, 122}

\bibitem[\protect\citeauthoryear{{Gaia Collaboration} et~al.,}{{Gaia Collaboration} et~al.}{2016}]{GaiaCollaboration2016}
{Gaia Collaboration} et~al., 2016, \mn@doi [\aap] {10.1051/0004-6361/201629272}, \href {https://ui.adsabs.harvard.edu/abs/2016A&A...595A...1G} {595, A1}

\bibitem[\protect\citeauthoryear{{Gaia Collaboration} et~al.,}{{Gaia Collaboration} et~al.}{2018}]{GaiaCollaboration2018}
{Gaia Collaboration} et~al., 2018, \mn@doi [\aap] {10.1051/0004-6361/201833051}, \href {https://ui.adsabs.harvard.edu/abs/2018A&A...616A...1G} {616, A1}

\bibitem[\protect\citeauthoryear{{George} \& {Stevens}}{{George} \& {Stevens}}{2007}]{george2007}
{George} S.~J.,  {Stevens} I.~R.,  2007, \mn@doi [\mnras] {10.1111/j.1365-2966.2007.12387.x}, \href {https://ui.adsabs.harvard.edu/abs/2007MNRAS.382..455G} {382, 455}

\bibitem[\protect\citeauthoryear{{Gizis} et~al.,}{{Gizis} et~al.}{2016}]{gizis2016}
{Gizis} J.~E.,  et~al., 2016, \mn@doi [\aj] {10.3847/0004-6256/152/5/123}, \href {https://ui.adsabs.harvard.edu/abs/2016AJ....152..123G} {152, 123}

\bibitem[\protect\citeauthoryear{{Gray}, {Corbally}, {Garrison}, {McFadden}, {Bubar}, {McGahee}, {O'Donoghue}  \& {Knox}}{{Gray} et~al.}{2006}]{gray2006}
{Gray} R.~O.,  {Corbally} C.~J.,  {Garrison} R.~F.,  {McFadden} M.~T.,  {Bubar} E.~J.,  {McGahee} C.~E.,  {O'Donoghue} A.~A.,   {Knox} E.~R.,  2006, \mn@doi [\aj] {10.1086/504637}, \href {https://ui.adsabs.harvard.edu/abs/2006AJ....132..161G} {132, 161}

\bibitem[\protect\citeauthoryear{{Green} \& {Madhusudhan}}{{Green} \& {Madhusudhan}}{2021}]{green2021}
{Green} D.~A.,  {Madhusudhan} N.,  2021, \mn@doi [\mnras] {10.1093/mnras/staa3208}, \href {https://ui.adsabs.harvard.edu/abs/2021MNRAS.500..211G} {500, 211}

\bibitem[\protect\citeauthoryear{{Griessmeier}}{{Griessmeier}}{2017}]{griessmeier2017}
{Griessmeier} J.~M.,  2017, in {Fischer} G.,  {Mann} G.,  {Panchenko} M.,   {Zarka} P.,  eds, Planetary Radio Emissions VIII. pp 285--299, \mn@doi{10.1553/PRE8s285}

\bibitem[\protect\citeauthoryear{{Grie{\ss}meier} et~al.,}{{Grie{\ss}meier} et~al.}{2004}]{griessmeier2004}
{Grie{\ss}meier} J.~M.,  et~al., 2004, \mn@doi [\aap] {10.1051/0004-6361:20035684}, \href {https://ui.adsabs.harvard.edu/abs/2004A&A...425..753G} {425, 753}

\bibitem[\protect\citeauthoryear{{Grie{\ss}meier}, {Motschmann}, {Mann}  \& {Rucker}}{{Grie{\ss}meier} et~al.}{2005}]{griessmeier2005}
{Grie{\ss}meier} J.~M.,  {Motschmann} U.,  {Mann} G.,   {Rucker} H.~O.,  2005, \mn@doi [\aap] {10.1051/0004-6361:20041976}, \href {https://ui.adsabs.harvard.edu/abs/2005A&A...437..717G} {437, 717}

\bibitem[\protect\citeauthoryear{{Grie{\ss}meier}, {Zarka}  \& {Spreeuw}}{{Grie{\ss}meier} et~al.}{2007}]{griessmeier2007}
{Grie{\ss}meier} J.~M.,  {Zarka} P.,   {Spreeuw} H.,  2007, \mn@doi [\aap] {10.1051/0004-6361:20077397}, \href {https://ui.adsabs.harvard.edu/abs/2007A&A...475..359G} {475, 359}

\bibitem[\protect\citeauthoryear{{Guirado}, {Azulay}, {Gauza}, {P{\'e}rez-Torres}, {Rebolo}, {Climent}  \& {Zapatero Osorio}}{{Guirado} et~al.}{2018}]{guirado2018}
{Guirado} J.~C.,  {Azulay} R.,  {Gauza} B.,  {P{\'e}rez-Torres} M.~A.,  {Rebolo} R.,  {Climent} J.~B.,   {Zapatero Osorio} M.~R.,  2018, \mn@doi [\aap] {10.1051/0004-6361/201732130}, \href {https://ui.adsabs.harvard.edu/abs/2018A&A...610A..23G} {610, A23}

\bibitem[\protect\citeauthoryear{{Gupta} et~al.,}{{Gupta} et~al.}{2017}]{gupta2017}
{Gupta} Y.,  et~al., 2017, \mn@doi [Current Science] {10.18520/cs/v113/i04/707-714}, \href {https://ui.adsabs.harvard.edu/abs/2017CSci..113..707G} {113, 707}

\bibitem[\protect\citeauthoryear{{Hallinan} et~al.,}{{Hallinan} et~al.}{2007}]{hallinan2007}
{Hallinan} G.,  et~al., 2007, \mn@doi [\apjl] {10.1086/519790}, \href {https://ui.adsabs.harvard.edu/abs/2007ApJ...663L..25H} {663, L25}

\bibitem[\protect\citeauthoryear{{Hallinan}, {Antonova}, {Doyle}, {Bourke}, {Lane}  \& {Golden}}{{Hallinan} et~al.}{2008}]{hallinan2008}
{Hallinan} G.,  {Antonova} A.,  {Doyle} J.~G.,  {Bourke} S.,  {Lane} C.,   {Golden} A.,  2008, \mn@doi [\apj] {10.1086/590360}, \href {https://ui.adsabs.harvard.edu/abs/2008ApJ...684..644H} {684, 644}

\bibitem[\protect\citeauthoryear{{Hallinan}, {Sirothia}, {Antonova}, {Ishwara-Chand ra}, {Bourke}, {Doyle}, {Hartman}  \& {Golden}}{{Hallinan} et~al.}{2013}]{hallinan2013}
{Hallinan} G.,  {Sirothia} S.~K.,  {Antonova} A.,  {Ishwara-Chand ra} C.~H.,  {Bourke} S.,  {Doyle} J.~G.,  {Hartman} J.,   {Golden} A.,  2013, \mn@doi [\apj] {10.1088/0004-637X/762/1/34}, \href {https://ui.adsabs.harvard.edu/abs/2013ApJ...762...34H} {762, 34}

\bibitem[\protect\citeauthoryear{{Hallinan} et~al.,}{{Hallinan} et~al.}{2015}]{hallinan2015}
{Hallinan} G.,  et~al., 2015, \mn@doi [\nat] {10.1038/nature14619}, \href {https://ui.adsabs.harvard.edu/abs/2015Natur.523..568H} {523, 568}

\bibitem[\protect\citeauthoryear{{Hasegawa}, {Kanagawa}  \& {Turner}}{{Hasegawa} et~al.}{2021}]{hasegawa2021}
{Hasegawa} Y.,  {Kanagawa} K.~D.,   {Turner} N.~J.,  2021, \mn@doi [\apj] {10.3847/1538-4357/ac257b}, \href {https://ui.adsabs.harvard.edu/abs/2021ApJ...923...27H} {923, 27}

\bibitem[\protect\citeauthoryear{{Hess} \& {Zarka}}{{Hess} \& {Zarka}}{2011}]{hess2011}
{Hess} S.~L.~G.,  {Zarka} P.,  2011, \mn@doi [\aap] {10.1051/0004-6361/201116510}, \href {https://ui.adsabs.harvard.edu/abs/2011A&A...531A..29H} {531, A29}

\bibitem[\protect\citeauthoryear{{Hill}}{{Hill}}{1979}]{hill1979}
{Hill} T.~W.,  1979, \mn@doi [\jgr] {10.1029/JA084iA11p06554}, \href {https://ui.adsabs.harvard.edu/abs/1979JGR....84.6554H} {84, 6554}

\bibitem[\protect\citeauthoryear{{Hill}}{{Hill}}{2001}]{hill2001}
{Hill} T.~W.,  2001, \mn@doi [\jgr] {10.1029/2000JA000302}, \href {https://ui.adsabs.harvard.edu/abs/2001JGR...106.8101H} {106, 8101}

\bibitem[\protect\citeauthoryear{{Hughes}, {Boley}, {Osten}, {White}  \& {Leacock}}{{Hughes} et~al.}{2021}]{hughes2021}
{Hughes} A.~G.,  {Boley} A.~C.,  {Osten} R.~A.,  {White} J.~A.,   {Leacock} M.,  2021, \mn@doi [\aj] {10.3847/1538-3881/ac02c3}, \href {https://ui.adsabs.harvard.edu/abs/2021AJ....162...43H} {162, 43}

\bibitem[\protect\citeauthoryear{{Kale} \& {Ishwara-Chandra}}{{Kale} \& {Ishwara-Chandra}}{2021}]{Kale2021}
{Kale} R.,  {Ishwara-Chandra} C.~H.,  2021, \mn@doi [Experimental Astronomy] {10.1007/s10686-020-09677-6}, \href {https://ui.adsabs.harvard.edu/abs/2021ExA....51...95K} {51, 95}

\bibitem[\protect\citeauthoryear{{Kao} \& {Sebastian Pineda}}{{Kao} \& {Sebastian Pineda}}{2022}]{kao2022}
{Kao} M.~M.,  {Sebastian Pineda} J.,  2022, \mn@doi [\apj] {10.3847/1538-4357/ac660b}, \href {https://ui.adsabs.harvard.edu/abs/2022ApJ...932...21K} {932, 21}

\bibitem[\protect\citeauthoryear{{Kao}, {Hallinan}, {Pineda}, {Escala}, {Burgasser}, {Bourke}  \& {Stevenson}}{{Kao} et~al.}{2016}]{kao2016}
{Kao} M.~M.,  {Hallinan} G.,  {Pineda} J.~S.,  {Escala} I.,  {Burgasser} A.,  {Bourke} S.,   {Stevenson} D.,  2016, \mn@doi [\apj] {10.3847/0004-637X/818/1/24}, \href {https://ui.adsabs.harvard.edu/abs/2016ApJ...818...24K} {818, 24}

\bibitem[\protect\citeauthoryear{{Kao}, {Hallinan}, {Pineda}, {Stevenson}  \& {Burgasser}}{{Kao} et~al.}{2018}]{kao2018}
{Kao} M.~M.,  {Hallinan} G.,  {Pineda} J.~S.,  {Stevenson} D.,   {Burgasser} A.,  2018, \mn@doi [\apjs] {10.3847/1538-4365/aac2d5}, \href {https://ui.adsabs.harvard.edu/abs/2018ApJS..237...25K} {237, 25}

\bibitem[\protect\citeauthoryear{{Katarzy{\'n}ski}, {Gawro{\'n}ski}  \& {Go{\'z}dziewski}}{{Katarzy{\'n}ski} et~al.}{2016}]{katarzynski2016}
{Katarzy{\'n}ski} K.,  {Gawro{\'n}ski} M.,   {Go{\'z}dziewski} K.,  2016, \mn@doi [\mnras] {10.1093/mnras/stw1354}, \href {https://ui.adsabs.harvard.edu/abs/2016MNRAS.461..929K} {461, 929}

\bibitem[\protect\citeauthoryear{{Kimura} et~al.,}{{Kimura} et~al.}{2013}]{kimura2013}
{Kimura} T.,  et~al., 2013, \mn@doi [Journal of Geophysical Research (Space Physics)] {10.1002/2013JA018833}, \href {https://ui.adsabs.harvard.edu/abs/2013JGRA..118.7019K} {118, 7019}

\bibitem[\protect\citeauthoryear{{Lacour} et~al.,}{{Lacour} et~al.}{2021}]{lacour2021}
{Lacour} S.,  et~al., 2021, \mn@doi [\aap] {10.1051/0004-6361/202141889}, \href {https://ui.adsabs.harvard.edu/abs/2021A&A...654L...2L} {654, L2}

\bibitem[\protect\citeauthoryear{{Ladreiter} \& {Leblanc}}{{Ladreiter} \& {Leblanc}}{1989}]{ladreiter1989}
{Ladreiter} H.~P.,  {Leblanc} Y.,  1989, \aap, \href {https://ui.adsabs.harvard.edu/abs/1989A&A...226..297L} {226, 297}

\bibitem[\protect\citeauthoryear{{Lazio} \& {Farrell}}{{Lazio} \& {Farrell}}{2007}]{lazio2007}
{Lazio} T. J.~W.,  {Farrell} W.~M.,  2007, \mn@doi [\apj] {10.1086/519730}, \href {https://ui.adsabs.harvard.edu/abs/2007ApJ...668.1182L} {668, 1182}

\bibitem[\protect\citeauthoryear{{Lazio}, {Farrell}, {Dietrick}, {Greenlees}, {Hogan}, {Jones}  \& {Hennig}}{{Lazio} et~al.}{2004}]{lazio2004}
{Lazio} T.~Joseph W.,  {Farrell} W.~M.,  {Dietrick} J.,  {Greenlees} E.,  {Hogan} E.,  {Jones} C.,   {Hennig} L.~A.,  2004, \mn@doi [\apj] {10.1086/422449}, \href {https://ui.adsabs.harvard.edu/abs/2004ApJ...612..511L} {612, 511}

\bibitem[\protect\citeauthoryear{{Lazio}, {Carmichael}, {Clark}, {Elkins}, {Gudmundsen}, {Mott}, {Szwajkowski}  \& {Hennig}}{{Lazio} et~al.}{2010a}]{lazio2010a}
{Lazio} T. J.~W.,  {Carmichael} S.,  {Clark} J.,  {Elkins} E.,  {Gudmundsen} P.,  {Mott} Z.,  {Szwajkowski} M.,   {Hennig} L.~A.,  2010a, \mn@doi [\aj] {10.1088/0004-6256/139/1/96}, \href {https://ui.adsabs.harvard.edu/abs/2010AJ....139...96L} {139, 96}

\bibitem[\protect\citeauthoryear{{Lazio}, {Shankland}, {Farrell}  \& {Blank}}{{Lazio} et~al.}{2010b}]{lazio2010b}
{Lazio} T. J.~W.,  {Shankland} P.~D.,  {Farrell} W.~M.,   {Blank} D.~L.,  2010b, \mn@doi [\aj] {10.1088/0004-6256/140/6/1929}, \href {https://ui.adsabs.harvard.edu/abs/2010AJ....140.1929L} {140, 1929}

\bibitem[\protect\citeauthoryear{{Lazio} et~al.,}{{Lazio} et~al.}{2019}]{lazio2019}
{Lazio} J.,  et~al., 2019, \mn@doi [\baas] {10.48550/arXiv.1803.06487}, \href {https://ui.adsabs.harvard.edu/abs/2019BAAS...51c.135L} {51, 135}

\bibitem[\protect\citeauthoryear{{Lecavelier Des Etangs}, {Sirothia}, {Gopal-Krishna}  \& {Zarka}}{{Lecavelier Des Etangs} et~al.}{2009}]{lecavelier2009}
{Lecavelier Des Etangs} A.,  {Sirothia} S.~K.,  {Gopal-Krishna}  {Zarka} P.,  2009, \mn@doi [\aap] {10.1051/0004-6361/200912347}, \href {https://ui.adsabs.harvard.edu/abs/2009A&A...500L..51L} {500, L51}

\bibitem[\protect\citeauthoryear{{Lecavelier Des Etangs}, {Sirothia}, {Gopal-Krishna}  \& {Zarka}}{{Lecavelier Des Etangs} et~al.}{2011}]{lecavelier2011}
{Lecavelier Des Etangs} A.,  {Sirothia} S.~K.,  {Gopal-Krishna}  {Zarka} P.,  2011, \mn@doi [\aap] {10.1051/0004-6361/201117330}, \href {https://ui.adsabs.harvard.edu/abs/2011A&A...533A..50L} {533, A50}

\bibitem[\protect\citeauthoryear{{Lecavelier des Etangs}, {Sirothia}, {Gopal-Krishna}  \& {Zarka}}{{Lecavelier des Etangs} et~al.}{2013}]{lecavelier2013}
{Lecavelier des Etangs} A.,  {Sirothia} S.~K.,  {Gopal-Krishna}  {Zarka} P.,  2013, \mn@doi [\aap] {10.1051/0004-6361/201219789}, \href {https://ui.adsabs.harvard.edu/abs/2013A&A...552A..65L} {552, A65}

\bibitem[\protect\citeauthoryear{{Li} et~al.,}{{Li} et~al.}{2018}]{li2018}
{Li} L.,  et~al., 2018, \mn@doi [Nature Communications] {10.1038/s41467-018-06107-2}, \href {https://ui.adsabs.harvard.edu/abs/2018NatCo...9.3709L} {9, 3709}

\bibitem[\protect\citeauthoryear{{Lynch}, {Murphy}, {Ravi}, {Hobbs}, {Lo}  \& {Ward}}{{Lynch} et~al.}{2016}]{lynch2016}
{Lynch} C.,  {Murphy} T.,  {Ravi} V.,  {Hobbs} G.,  {Lo} K.,   {Ward} C.,  2016, \mn@doi [\mnras] {10.1093/mnras/stw050}, \href {https://ui.adsabs.harvard.edu/abs/2016MNRAS.457.1224L} {457, 1224}

\bibitem[\protect\citeauthoryear{{Lynch}, {Murphy}, {Kaplan}, {Ireland}  \& {Bell}}{{Lynch} et~al.}{2017}]{lynch2017}
{Lynch} C.~R.,  {Murphy} T.,  {Kaplan} D.~L.,  {Ireland} M.,   {Bell} M.~E.,  2017, \mn@doi [\mnras] {10.1093/mnras/stx354}, \href {https://ui.adsabs.harvard.edu/abs/2017MNRAS.467.3447L} {467, 3447}

\bibitem[\protect\citeauthoryear{{Lynch}, {Murphy}, {Lenc}  \& {Kaplan}}{{Lynch} et~al.}{2018}]{lynch2018}
{Lynch} C.~R.,  {Murphy} T.,  {Lenc} E.,   {Kaplan} D.~L.,  2018, \mn@doi [\mnras] {10.1093/mnras/sty1138}, \href {https://ui.adsabs.harvard.edu/abs/2018MNRAS.478.1763L} {478, 1763}

\bibitem[\protect\citeauthoryear{{Majid}, {Winterhalter}, {Chandra}, {Kuiper}, {Lazio}, {Naudet}  \& {Zarka}}{{Majid} et~al.}{2006}]{majid2006}
{Majid} W.,  {Winterhalter} D.,  {Chandra} I.,  {Kuiper} T.,  {Lazio} J.,  {Naudet} C.,   {Zarka} P.,  2006, in Planetary Radio Emissions VI. pp 589--594

\bibitem[\protect\citeauthoryear{{Mamajek} \& {Bell}}{{Mamajek} \& {Bell}}{2014}]{mamajek2014}
{Mamajek} E.~E.,  {Bell} C. P.~M.,  2014, \mn@doi [\mnras] {10.1093/mnras/stu1894}, \href {https://ui.adsabs.harvard.edu/abs/2014MNRAS.445.2169M} {445, 2169}

\bibitem[\protect\citeauthoryear{{McLean}, {Berger}, {Irwin}, {Forbrich}  \& {Reiners}}{{McLean} et~al.}{2011}]{mclean2011}
{McLean} M.,  {Berger} E.,  {Irwin} J.,  {Forbrich} J.,   {Reiners} A.,  2011, \mn@doi [\apj] {10.1088/0004-637X/741/1/27}, \href {https://ui.adsabs.harvard.edu/abs/2011ApJ...741...27M} {741, 27}

\bibitem[\protect\citeauthoryear{{McLean}, {Berger}  \& {Reiners}}{{McLean} et~al.}{2012}]{mclean2012}
{McLean} M.,  {Berger} E.,   {Reiners} A.,  2012, \mn@doi [\apj] {10.1088/0004-637X/746/1/23}, \href {https://ui.adsabs.harvard.edu/abs/2012ApJ...746...23M} {746, 23}

\bibitem[\protect\citeauthoryear{{Murphy} et~al.,}{{Murphy} et~al.}{2015}]{Murphy2015}
{Murphy} T.,  et~al., 2015, \mn@doi [\mnras] {10.1093/mnras/stu2253}, \href {https://ui.adsabs.harvard.edu/abs/2015MNRAS.446.2560M} {446, 2560}

\bibitem[\protect\citeauthoryear{{Narang}}{{Narang}}{2022}]{narang2022}
{Narang} M.,  2022, \mn@doi [\mnras] {10.1093/mnras/stac1902}, \href {https://ui.adsabs.harvard.edu/abs/2022MNRAS.515.2015N} {515, 2015}

\bibitem[\protect\citeauthoryear{{Narang}, {Manoj}  \& {Ishwara Chandra}}{{Narang} et~al.}{2021a}]{narang2021b}
{Narang} M.,  {Manoj} P.,   {Ishwara Chandra} C.~H.,  2021a, \mn@doi [Research Notes of the American Astronomical Society] {10.3847/2515-5172/ac0fe0}, \href {https://ui.adsabs.harvard.edu/abs/2021RNAAS...5..158N} {5, 158}

\bibitem[\protect\citeauthoryear{{Narang} et~al.,}{{Narang} et~al.}{2021b}]{narang2021a}
{Narang} M.,  et~al., 2021b, \mn@doi [\mnras] {10.1093/mnras/staa3565}, \href {https://ui.adsabs.harvard.edu/abs/2021MNRAS.500.4818N} {500, 4818}

\bibitem[\protect\citeauthoryear{{Narang}, {Oza}, {Hakim}, {Manoj}, {Banyal}  \& {Thorngren}}{{Narang} et~al.}{2023a}]{narang2023a}
{Narang} M.,  {Oza} A.~V.,  {Hakim} K.,  {Manoj} P.,  {Banyal} R.~K.,   {Thorngren} D.~P.,  2023a, \mn@doi [\aj] {10.3847/1538-3881/ac9eb8}, \href {https://ui.adsabs.harvard.edu/abs/2023AJ....165....1N} {165, 1}

\bibitem[\protect\citeauthoryear{{Narang} et~al.,}{{Narang} et~al.}{2023b}]{narang2023b}
{Narang} M.,  et~al., 2023b, \mn@doi [\mnras] {10.1093/mnras/stad1027}, \href {https://ui.adsabs.harvard.edu/abs/2023MNRAS.522.1662N} {522, 1662}

\bibitem[\protect\citeauthoryear{{Nichols}}{{Nichols}}{2011}]{nichols2011}
{Nichols} J.~D.,  2011, in EPSC-DPS Joint Meeting 2011. p.~1353

\bibitem[\protect\citeauthoryear{{Nichols}}{{Nichols}}{2012}]{nichols2012}
{Nichols} J.~D.,  2012, \mn@doi [\mnras] {10.1111/j.1745-3933.2012.01348.x}, \href {https://ui.adsabs.harvard.edu/abs/2012MNRAS.427L..75N} {427, L75}

\bibitem[\protect\citeauthoryear{{Nielsen} et~al.,}{{Nielsen} et~al.}{2020}]{nielsen2020}
{Nielsen} E.~L.,  et~al., 2020, \mn@doi [\aj] {10.3847/1538-3881/ab5b92}, \href {https://ui.adsabs.harvard.edu/abs/2020AJ....159...71N} {159, 71}

\bibitem[\protect\citeauthoryear{{Nowak} et~al.,}{{Nowak} et~al.}{2020}]{nowak2020}
{Nowak} M.,  et~al., 2020, \mn@doi [\aap] {10.1051/0004-6361/202039039}, \href {https://ui.adsabs.harvard.edu/abs/2020A&A...642L...2N} {642, L2}

\bibitem[\protect\citeauthoryear{{Noyola}, {Satyal}  \& {Musielak}}{{Noyola} et~al.}{2014}]{noyola2014}
{Noyola} J.~P.,  {Satyal} S.,   {Musielak} Z.~E.,  2014, \mn@doi [\apj] {10.1088/0004-637X/791/1/25}, \href {https://ui.adsabs.harvard.edu/abs/2014ApJ...791...25N} {791, 25}

\bibitem[\protect\citeauthoryear{{Noyola}, {Satyal}  \& {Musielak}}{{Noyola} et~al.}{2016}]{noyola2016}
{Noyola} J.~P.,  {Satyal} S.,   {Musielak} Z.~E.,  2016, \mn@doi [\apj] {10.3847/0004-637X/821/2/97}, \href {https://ui.adsabs.harvard.edu/abs/2016ApJ...821...97N} {821, 97}

\bibitem[\protect\citeauthoryear{{O'Gorman}, {Coughlan}, {Vlemmings}, {Varenius}, {Sirothia}, {Ray}  \& {Olofsson}}{{O'Gorman} et~al.}{2018}]{o'gorman2018}
{O'Gorman} E.,  {Coughlan} C.~P.,  {Vlemmings} W.,  {Varenius} E.,  {Sirothia} S.,  {Ray} T.~P.,   {Olofsson} H.,  2018, \mn@doi [\aap] {10.1051/0004-6361/201731965}, \href {https://ui.adsabs.harvard.edu/abs/2018A&A...612A..52O} {612, A52}

\bibitem[\protect\citeauthoryear{{Osten}, {Hawley}, {Bastian}  \& {Reid}}{{Osten} et~al.}{2006}]{osten2006}
{Osten} R.~A.,  {Hawley} S.~L.,  {Bastian} T.~S.,   {Reid} I.~N.,  2006, \mn@doi [\apj] {10.1086/498345}, \href {https://ui.adsabs.harvard.edu/abs/2006ApJ...637..518O} {637, 518}

\bibitem[\protect\citeauthoryear{{Osten}, {Phan-Bao}, {Hawley}, {Reid}  \& {Ojha}}{{Osten} et~al.}{2009}]{osten2009}
{Osten} R.~A.,  {Phan-Bao} N.,  {Hawley} S.~L.,  {Reid} I.~N.,   {Ojha} R.,  2009, \mn@doi [\apj] {10.1088/0004-637X/700/2/1750}, \href {https://ui.adsabs.harvard.edu/abs/2009ApJ...700.1750O} {700, 1750}

\bibitem[\protect\citeauthoryear{{P{\'e}rez-Torres} et~al.,}{{P{\'e}rez-Torres} et~al.}{2021}]{perez-torres2021}
{P{\'e}rez-Torres} M.,  et~al., 2021, \mn@doi [\aap] {10.1051/0004-6361/202039052}, \href {https://ui.adsabs.harvard.edu/abs/2021A&A...645A..77P} {645, A77}

\bibitem[\protect\citeauthoryear{{Phan-Bao}, {Osten}, {Lim}, {Mart{\'\i}n}  \& {Ho}}{{Phan-Bao} et~al.}{2007}]{phan-bao2007}
{Phan-Bao} N.,  {Osten} R.~A.,  {Lim} J.,  {Mart{\'\i}n} E.~L.,   {Ho} P. T.~P.,  2007, \mn@doi [\apj] {10.1086/511061}, \href {https://ui.adsabs.harvard.edu/abs/2007ApJ...658..553P} {658, 553}

\bibitem[\protect\citeauthoryear{{Pineda} \& {Villadsen}}{{Pineda} \& {Villadsen}}{2023}]{pineda2023}
{Pineda} J.~S.,  {Villadsen} J.,  2023, \mn@doi [Nature Astronomy] {10.1038/s41550-023-01914-0}, \href {https://ui.adsabs.harvard.edu/abs/2023NatAs...7..569P} {7, 569}

\bibitem[\protect\citeauthoryear{{Reiners} \& {Christensen}}{{Reiners} \& {Christensen}}{2010}]{reiners2010}
{Reiners} A.,  {Christensen} U.~R.,  2010, \mn@doi [\aap] {10.1051/0004-6361/201014251}, \href {https://ui.adsabs.harvard.edu/abs/2010A&A...522A..13R} {522, A13}

\bibitem[\protect\citeauthoryear{{Richey-Yowell}, {Kao}, {Pineda}, {Shkolnik}  \& {Hallinan}}{{Richey-Yowell} et~al.}{2020}]{richey-yowell2020}
{Richey-Yowell} T.,  {Kao} M.~M.,  {Pineda} J.~S.,  {Shkolnik} E.~L.,   {Hallinan} G.,  2020, \mn@doi [\apj] {10.3847/1538-4357/abb826}, \href {https://ui.adsabs.harvard.edu/abs/2020ApJ...903...74R} {903, 74}

\bibitem[\protect\citeauthoryear{{Robinson} \& {Vondrak}}{{Robinson} \& {Vondrak}}{1984}]{robinson1984}
{Robinson} R.~M.,  {Vondrak} R.~R.,  1984, \mn@doi [\jgr] {10.1029/JA089iA06p03951}, \href {https://ui.adsabs.harvard.edu/abs/1984JGR....89.3951R} {89, 3951}

\bibitem[\protect\citeauthoryear{{Route} \& {Wolszczan}}{{Route} \& {Wolszczan}}{2016}]{route2016}
{Route} M.,  {Wolszczan} A.,  2016, \mn@doi [\apj] {10.3847/0004-637X/830/2/85}, \href {https://ui.adsabs.harvard.edu/abs/2016ApJ...830...85R} {830, 85}

\bibitem[\protect\citeauthoryear{{Russell}}{{Russell}}{1978}]{russell1978}
{Russell} C.~T.,  1978, \mn@doi [\nat] {10.1038/272147a0}, \href {https://ui.adsabs.harvard.edu/abs/1978Natur.272..147R} {272, 147}

\bibitem[\protect\citeauthoryear{{Ryabov}, {Zarka}  \& {Ryabov}}{{Ryabov} et~al.}{2004}]{ryabov2004}
{Ryabov} V.~B.,  {Zarka} P.,   {Ryabov} B.~P.,  2004, \mn@doi [\planss] {10.1016/j.pss.2004.09.019}, \href {https://ui.adsabs.harvard.edu/abs/2004P&SS...52.1479R} {52, 1479}

\bibitem[\protect\citeauthoryear{{S{\'a}nchez-Lavega}}{{S{\'a}nchez-Lavega}}{2004}]{sanchez-Lavega2004}
{S{\'a}nchez-Lavega} A.,  2004, \mn@doi [\apjl] {10.1086/422840}, \href {https://ui.adsabs.harvard.edu/abs/2004ApJ...609L..87S} {609, L87}

\bibitem[\protect\citeauthoryear{{Sanz-Forcada}, {Micela}, {Ribas}, {Pollock}, {Eiroa}, {Velasco}, {Solano}  \& {Garc{\'\i}a-{\'A}lvarez}}{{Sanz-Forcada} et~al.}{2011}]{sanz-forcada2011}
{Sanz-Forcada} J.,  {Micela} G.,  {Ribas} I.,  {Pollock} A.~M.~T.,  {Eiroa} C.,  {Velasco} A.,  {Solano} E.,   {Garc{\'\i}a-{\'A}lvarez} D.,  2011, \mn@doi [\aap] {10.1051/0004-6361/201116594}, \href {https://ui.adsabs.harvard.edu/abs/2011A&A...532A...6S} {532, A6}

\bibitem[\protect\citeauthoryear{{Saur}, {Grambusch}, {Duling}, {Neubauer}  \& {Simon}}{{Saur} et~al.}{2013}]{saur2013}
{Saur} J.,  {Grambusch} T.,  {Duling} S.,  {Neubauer} F.~M.,   {Simon} S.,  2013, \mn@doi [\aap] {10.1051/0004-6361/201118179}, \href {https://ui.adsabs.harvard.edu/abs/2013A&A...552A.119S} {552, A119}

\bibitem[\protect\citeauthoryear{{Shiratori}, {Yokoo}, {Saso}, {Kameya}, {Iwadate}  \& {Asari}}{{Shiratori} et~al.}{2006}]{shiratori2006}
{Shiratori} Y.,  {Yokoo} H.,  {Saso} T.,  {Kameya} O.,  {Iwadate} K.,   {Asari} K.,  2006, in {Arnold} L.,  {Bouchy} F.,   {Moutou} C.,  eds, Tenth Anniversary of 51 Peg-b: Status of and prospects for hot Jupiter studies. pp 290--292

\bibitem[\protect\citeauthoryear{{Smith}, {Collier Cameron}, {Greaves}, {Jardine}, {Langston}  \& {Backer}}{{Smith} et~al.}{2009}]{smith2009}
{Smith} A.~M.~S.,  {Collier Cameron} A.,  {Greaves} J.,  {Jardine} M.,  {Langston} G.,   {Backer} D.,  2009, \mn@doi [\mnras] {10.1111/j.1365-2966.2009.14510.x}, \href {https://ui.adsabs.harvard.edu/abs/2009MNRAS.395..335S} {395, 335}

\bibitem[\protect\citeauthoryear{{Snellen}, {Brandl}, {de Kok}, {Brogi}, {Birkby}  \& {Schwarz}}{{Snellen} et~al.}{2014}]{snellen2014}
{Snellen} I. A.~G.,  {Brandl} B.~R.,  {de Kok} R.~J.,  {Brogi} M.,  {Birkby} J.,   {Schwarz} H.,  2014, \mn@doi [\nat] {10.1038/nature13253}, \href {https://ui.adsabs.harvard.edu/abs/2014Natur.509...63S} {509, 63}

\bibitem[\protect\citeauthoryear{{Stevens}}{{Stevens}}{2005}]{stevens2005}
{Stevens} I.~R.,  2005, \mn@doi [\mnras] {10.1111/j.1365-2966.2004.08528.x}, \href {https://ui.adsabs.harvard.edu/abs/2005MNRAS.356.1053S} {356, 1053}

\bibitem[\protect\citeauthoryear{{Stroe}, {Snellen}  \& {R{\"o}ttgering}}{{Stroe} et~al.}{2012}]{stroe2012}
{Stroe} A.,  {Snellen} I.~A.~G.,   {R{\"o}ttgering} H.~J.~A.,  2012, \mn@doi [\aap] {10.1051/0004-6361/201220006}, \href {https://ui.adsabs.harvard.edu/abs/2012A&A...546A.116S} {546, A116}

\bibitem[\protect\citeauthoryear{{Tao}, {Fujiwara}  \& {Kasaba}}{{Tao} et~al.}{2010}]{tao2010}
{Tao} C.,  {Fujiwara} H.,   {Kasaba} Y.,  2010, \mn@doi [\planss] {10.1016/j.pss.2009.10.005}, \href {https://ui.adsabs.harvard.edu/abs/2010P&SS...58..351T} {58, 351}

\bibitem[\protect\citeauthoryear{Treumann}{Treumann}{2006}]{treumann2006}
Treumann R.~A.,  2006, The Astronomy and Astrophysics Review, 13, 229

\bibitem[\protect\citeauthoryear{{Turner}, {Griessmeier}, {Zarka}  \& {Vasylieva}}{{Turner} et~al.}{2017}]{turner2017}
{Turner} J.~D.,  {Griessmeier} J.~M.,  {Zarka} P.,   {Vasylieva} I.,  2017, in {Fischer} G.,  {Mann} G.,  {Panchenko} M.,   {Zarka} P.,  eds, Planetary Radio Emissions VIII. pp 301--313 (\mn@eprint {arXiv} {1710.04997}), \mn@doi{10.1553/PRE8s301}

\bibitem[\protect\citeauthoryear{{Turner} et~al.,}{{Turner} et~al.}{2021}]{turner2021}
{Turner} J.~D.,  et~al., 2021, \mn@doi [\aap] {10.1051/0004-6361/201937201}, \href {https://ui.adsabs.harvard.edu/abs/2021A&A...645A..59T} {645, A59}

\bibitem[\protect\citeauthoryear{{Turnpenney}, {Nichols}, {Wynn}  \& {Burleigh}}{{Turnpenney} et~al.}{2018}]{turnpenney2018}
{Turnpenney} S.,  {Nichols} J.~D.,  {Wynn} G.~A.,   {Burleigh} M.~R.,  2018, \mn@doi [\apj] {10.3847/1538-4357/aaa59c10.48550/arXiv.1801.01324}, \href {https://ui.adsabs.harvard.edu/abs/2018ApJ...854...72T} {854, 72}

\bibitem[\protect\citeauthoryear{{Vedantham} et~al.,}{{Vedantham} et~al.}{2020}]{vedantham2020}
{Vedantham} H.~K.,  et~al., 2020, \mn@doi [Nature Astronomy] {10.1038/s41550-020-1011-9}, \href {https://ui.adsabs.harvard.edu/abs/2020NatAs...4..577V} {4, 577}

\bibitem[\protect\citeauthoryear{{Vedantham} et~al.,}{{Vedantham} et~al.}{2023}]{vedantham2023}
{Vedantham} H.~K.,  et~al., 2023, \mn@doi [arXiv e-prints] {10.48550/arXiv.2301.01003}, \href {https://ui.adsabs.harvard.edu/abs/2023arXiv230101003V} {p. arXiv:2301.01003}

\bibitem[\protect\citeauthoryear{{Weber} et~al.,}{{Weber} et~al.}{2017a}]{weber2017a}
{Weber} C.,  et~al., 2017a, in {Fischer} G.,  {Mann} G.,  {Panchenko} M.,   {Zarka} P.,  eds, Planetary Radio Emissions VIII. pp 317--329, \mn@doi{10.1553/PRE8s317}

\bibitem[\protect\citeauthoryear{{Weber} et~al.,}{{Weber} et~al.}{2017b}]{weber2017b}
{Weber} C.,  et~al., 2017b, \mn@doi [\mnras] {10.1093/mnras/stx1099}, \href {https://ui.adsabs.harvard.edu/abs/2017MNRAS.469.3505W} {469, 3505}

\bibitem[\protect\citeauthoryear{{Weber}, {Erkaev}, {Ivanov}, {Odert}, {Grie{\ss}meier}, {Fossati}, {Lammer}  \& {Rucker}}{{Weber} et~al.}{2018}]{weber2018}
{Weber} C.,  {Erkaev} N.~V.,  {Ivanov} V.~A.,  {Odert} P.,  {Grie{\ss}meier} J.~M.,  {Fossati} L.,  {Lammer} H.,   {Rucker} H.~O.,  2018, \mn@doi [\mnras] {10.1093/mnras/sty2079}, \href {https://ui.adsabs.harvard.edu/abs/2018MNRAS.480.3680W} {480, 3680}

\bibitem[\protect\citeauthoryear{{Williams}, {Berger}  \& {Zauderer}}{{Williams} et~al.}{2013}]{williams2013}
{Williams} P. K.~G.,  {Berger} E.,   {Zauderer} B.~A.,  2013, \mn@doi [\apjl] {10.1088/2041-8205/767/2/L30}, \href {https://ui.adsabs.harvard.edu/abs/2013ApJ...767L..30W} {767, L30}

\bibitem[\protect\citeauthoryear{{Winglee}, {Dulk}  \& {Bastian}}{{Winglee} et~al.}{1986}]{winglee1986}
{Winglee} R.~M.,  {Dulk} G.~A.,   {Bastian} T.~S.,  1986, \mn@doi [\apjl] {10.1086/184760}, \href {https://ui.adsabs.harvard.edu/abs/1986ApJ...309L..59W} {309, L59}

\bibitem[\protect\citeauthoryear{{Winterhalter} et~al.,}{{Winterhalter} et~al.}{2006}]{winterhalter2006}
{Winterhalter} D.,  et~al., 2006, in Planetary Radio Emissions VI. pp 595--602

\bibitem[\protect\citeauthoryear{Wu \& Lee}{Wu \& Lee}{1979}]{wulee1979}
Wu C.,  Lee L.,  1979, The Astrophysical Journal, 230, 621

\bibitem[\protect\citeauthoryear{{Yantis}, {Sullivan}  \& {Erickson}}{{Yantis} et~al.}{1977}]{yantis1977}
{Yantis} W.~F.,  {Sullivan} W.~T. I.,   {Erickson} W.~C.,  1977, in Bulletin of the American Astronomical Society. p.~453

\bibitem[\protect\citeauthoryear{{Zarka}}{{Zarka}}{1992}]{zarka1992}
{Zarka} P.,  1992, \mn@doi [Advances in Space Research] {10.1016/0273-1177(92)90383-9}, \href {https://ui.adsabs.harvard.edu/abs/1992AdSpR..12...99Z} {12, 99}

\bibitem[\protect\citeauthoryear{{Zarka}}{{Zarka}}{1998}]{zarka1998}
{Zarka} P.,  1998, \mn@doi [\jgr] {10.1029/98JE01323}, \href {https://ui.adsabs.harvard.edu/abs/1998JGR...10320159Z} {103, 20159}

\bibitem[\protect\citeauthoryear{{Zarka}}{{Zarka}}{2007}]{zarka2007}
{Zarka} P.,  2007, \mn@doi [\planss] {10.1016/j.pss.2006.05.045}, \href {https://ui.adsabs.harvard.edu/abs/2007P&SS...55..598Z} {55, 598}

\bibitem[\protect\citeauthoryear{{Zarka}}{{Zarka}}{2011}]{zarka2011}
{Zarka} P.,  2011, in {Rucker} H.~O.,  {Kurth} W.~S.,  {Louarn} P.,   {Fischer} G.,  eds, Planetary, Solar and Heliospheric Radio Emissions (PRE VII). pp 287--301

\bibitem[\protect\citeauthoryear{{Zarka} et~al.,}{{Zarka} et~al.}{1997}]{zarka1997}
{Zarka} P.,  et~al., 1997, in Planetary Radio Emission IV. pp 101--127

\bibitem[\protect\citeauthoryear{{Zarka}, {Treumann}, {Ryabov}  \& {Ryabov}}{{Zarka} et~al.}{2001}]{zarka2001}
{Zarka} P.,  {Treumann} R.~A.,  {Ryabov} B.~P.,   {Ryabov} V.~B.,  2001, \mn@doi [\apss] {10.1023/A:1012221527425}, \href {https://ui.adsabs.harvard.edu/abs/2001Ap&SS.277..293Z} {277, 293}

\bibitem[\protect\citeauthoryear{{de Gasperin}, {Lazio}  \& {Knapp}}{{de Gasperin} et~al.}{2020}]{deGasperin2020}
{de Gasperin} F.,  {Lazio} T.~J.~W.,   {Knapp} M.,  2020, \mn@doi [\aap] {10.1051/0004-6361/202038746}, \href {https://ui.adsabs.harvard.edu/abs/2020A&A...644A.157D} {644, A157}

\makeatother
\end{thebibliography}









\begin{appendix}
\section{Evaluation of the core density}

To estimate the density profile within exoplanets where clues are virtually absent, it is helpful to employ the assumption of a polytropic gas sphere, $P=K\rho^{1+(1/n)}$, where $P$ is Pressure, $K$ is a constant, $\rho$ is density and $n$ is the polytropic index \citep[e.g.,][]{sanchez-Lavega2004, griessmeier2007}. Here, we set the polytropic index $n = 1.5$ and numerically solved the Lane-Emden equation,
\begin{equation}
    \label{appendix:LEeq}
    \frac{1}{\xi} \frac{d}{d\xi} \left(\xi^2\frac{d\theta}{d\xi}\right)  = -\theta(\xi) ^n ,
\end{equation}
where $\xi$ and $\theta(\xi)$ are non-dimension radius and density, respectively, and they are defined using the radius $r$ and density $\rho$ of the sphere as follows:
\begin{equation}
    \label{appendix:def_xi_theta}
    \xi = \frac{r}{\alpha} \ , \ \theta^{n}=\frac{\rho}{\rho_\mathrm{c}} ,
\end{equation}
where $\alpha$ is a constant, and $\rho_\mathrm{c}$ is the density at the centre of sphere. Eq. \ref{appendix:LEeq} can be rewritten as two differential equations through the conversion $y_1 = \theta$ and $y_2 = \frac{d\theta}{d\xi} = \theta'$,
\begin{equation}
\label{appendix:LEeq2}
\left\{ \,
 \begin{aligned}
    & \frac{dy_1}{d\xi} = y_2 \\
    & \frac{dy_2}{d\xi} = -\frac{2}{\xi}y_2 - y_1^{n} \ \ ,
 \end{aligned}
\right.
\end{equation}
and we can then obtain the functions of $\theta$ and $\theta'$ by solving Eq. \ref{appendix:LEeq2} with the initial conditions $\theta(0) = 1$ and $ \theta'(0) = 0$. From the solutions, a mass $M$ and a mean density $\bar{\rho}$ within an arbitrary distance $r$ from the centre can be determined:
\begin{equation}
    \label{appendix:mass}
    M = \int_{0}^{r} 4\pi r^2 \rho(r) dr = -4\pi \alpha^3 \rho_\mathrm{c} \left(\xi^2 \frac{d\theta}{d\xi}\right) .
\end{equation}
\begin{equation}
    \label{appendix:density}
    \bar{\rho} = \frac{M}{\frac{4}{3}\pi r^3} = -\frac{3\rho_\mathrm{c}}{\xi}\left(\frac{d\theta}{d\xi}\right) .
\end{equation}
Then, the density normalized by the mean density of the entire sphere can be computed for each $\xi$ applying the definition \ref{appendix:def_xi_theta},
\begin{equation}
    \label{appendix:appendix:density2}
    \frac{\rho}{\bar{\rho}} = \frac{\rho}{\rho_\mathrm{c}}\frac{\rho_\mathrm{c}}{\bar{\rho}}= -\frac{\xi\theta^n}{3}\left(\frac{d\theta}{d\xi}\right)^{-1} .
\end{equation}
When each $\xi$ is divided by $\xi=R/\alpha$, which is found by searching for the value of $\xi$ such that $\theta=0$, $\xi$ also can be expressed in terms of the radius normalized by the sphere's radius $r/R$.

\begin{figure*}
    \begin{tabular}{cc}
        \begin{minipage}[t]{0.45\hsize}
            \centering
            \includegraphics[width=0.75\hsize]{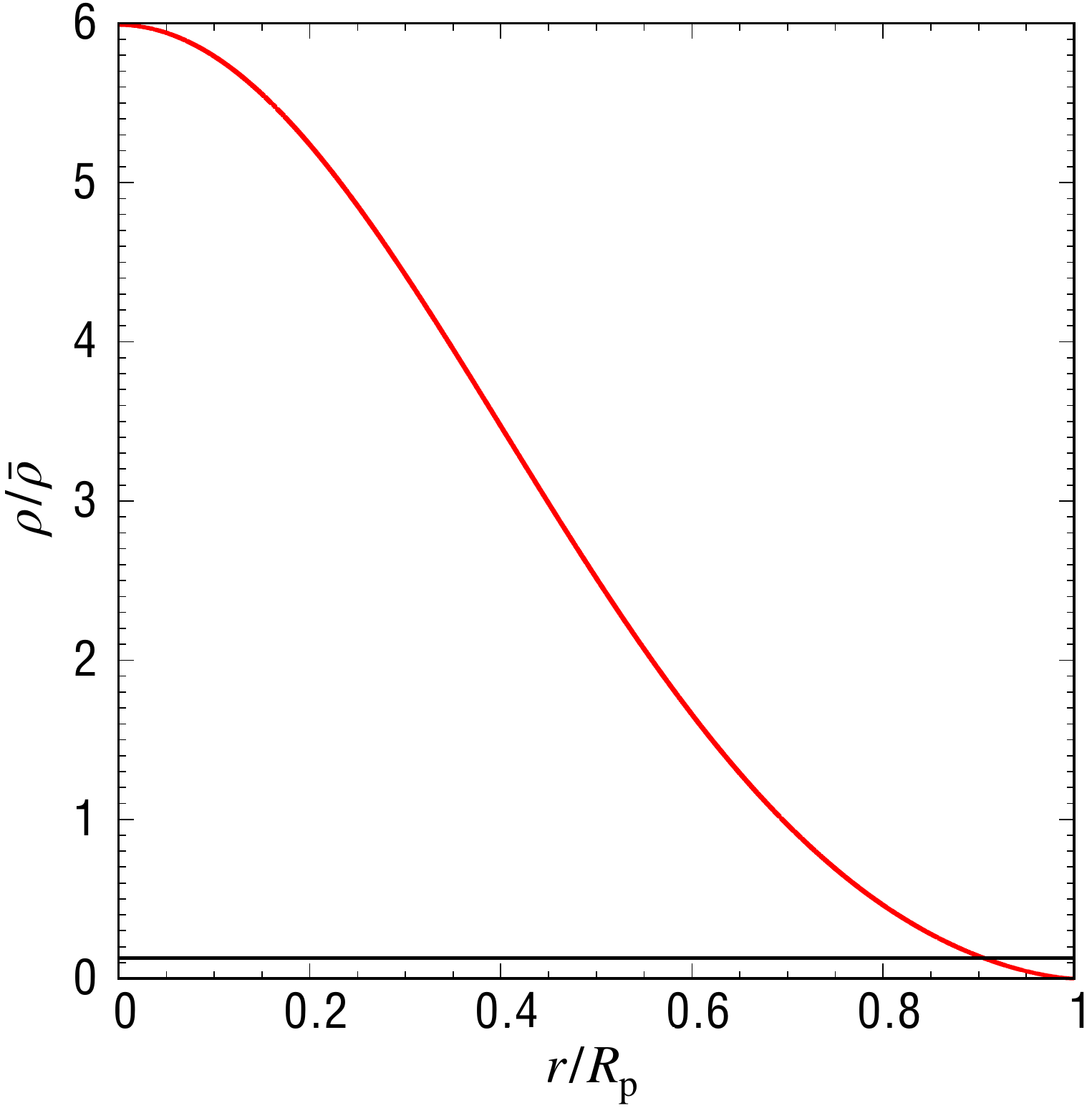}
        \end{minipage} &
        \begin{minipage}[t]{0.45\hsize}
            \centering
            \includegraphics[width=0.788\hsize]{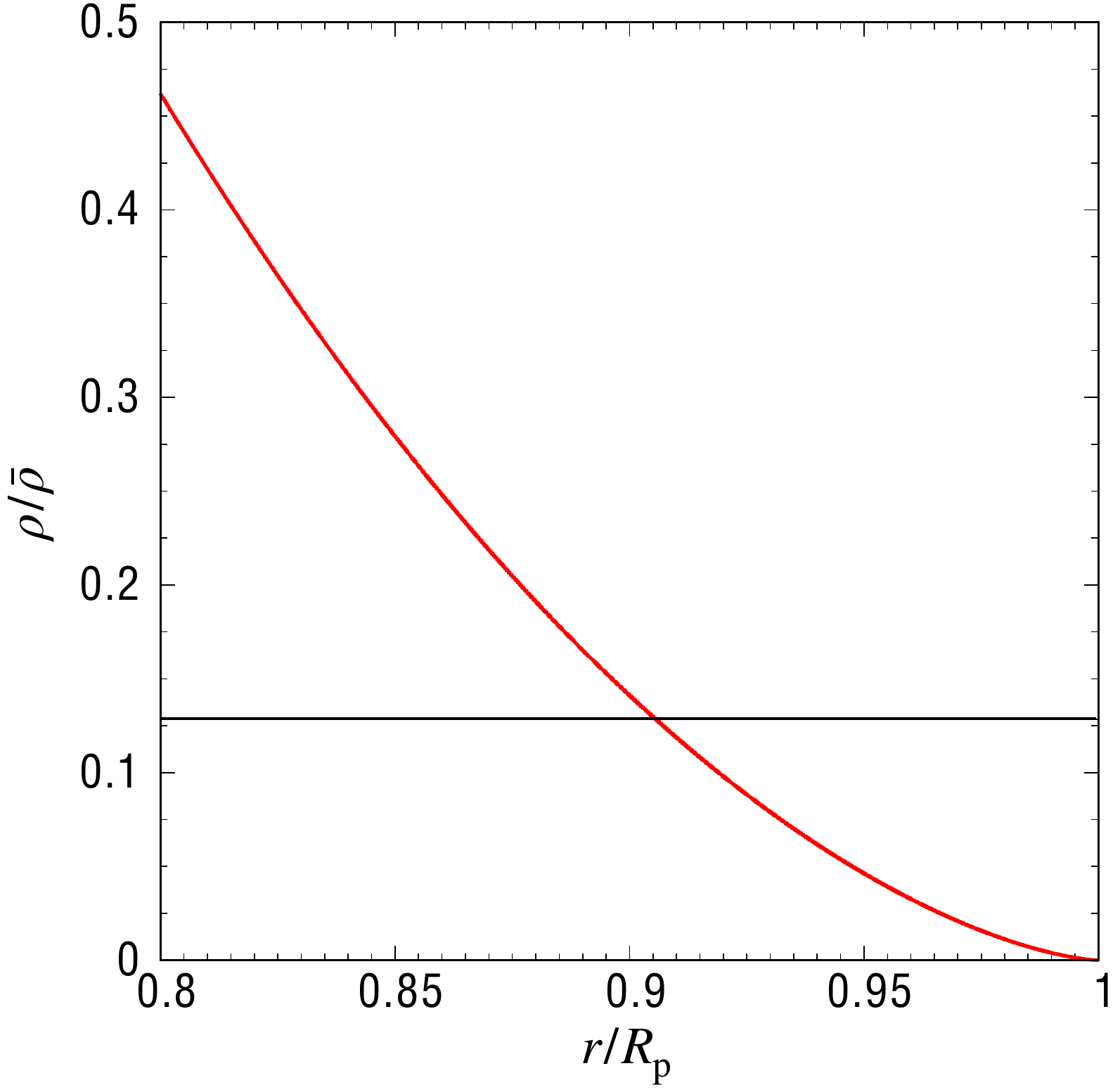}
        \end{minipage}\\
    \end{tabular}
    \caption{The estimation of the density profile of $\beta$ Pic b. The vertical axis represents density, and the horizontal axis represents distance from the center. Each axis is normalized by the mean density and the radius of $\beta$ Pic b, respectively. The red solid line shows the density profile. The black solid line corresponds to the density that hydrogen undergoes a phase transition to metallic, $\sim 0.7 \mathrm{g/cm^3}$. The right panel is an expanded version of the left panel.}
    \label{appendix:density_profile}
\end{figure*}

Based on the above, we can estimate the density profile inside an arbitrary volume by substituting the object's parameters. Figure \ref{appendix:density_profile} displays the estimated density profile of $\beta$ Pic b. The core radius is found from the point where the solid red line intersects the solid black line in Fig. \ref{appendix:density_profile}, $r_\mathrm{c,p} \sim 0.9 R_\mathrm{p}$. The core mass can computed by \ref{appendix:mass}, therefore, we can estimate the core density, $\rho_\mathrm{c,p} \sim 7.254\ \mathrm{g/cm^3}$. 
\end{appendix}

\bsp	
\label{lastpage}
\end{document}